# Parahydrogen-induced polarization relayed via proton exchange.

Kolja Them*,[1,6], Frowin Ellermann[1], Andrey N. Pravdivtsev[1], Oleg G. Salnikov[2,3,4], Ivan V. Skovpin[2,3], Igor V. Koptyug[2,3], Rainer Herges[5] and Jan-Bernd Hövener*,[1,7]

[1]Section Biomedical Imaging, Molecular Imaging North Competence Center (MOIN CC), Department of Radiology and Neuroradiology, University Medical Center Schleswig-Holstein and Kiel University, Am Botanischen Garten 14, 24118 Kiel, Germany

[2]International Tomography Center, SB RAS, 3A Institutskaya st., Novosibirsk 630090, Russia

[3]Novosibirsk State University, 2 Pirogova st., Novosibirsk 630090, Russia

[4]Boreskov Institute of Catalysis SB RAS, 4 Acad. Lavrentiev pr., Novosibirsk 630090, Russia

[5]Otto-Diels-Institute for Organic Chemistry, Christian-Albrechts-University, 24118 Kiel, Germany

[6]kolja.them@rad.uni-kiel.de, [7]jan.hoevener@rad.uni-kiel.de

**Abstract:** The sensitivity of NMR and MRI can be boosted via hyperpolarization of nuclear spins. However, current methods are costly, polarization is relatively low, or applicability is limited. Here, we report a new hyperpolarization method combining the low-cost, high polarization of hydrogenative parahydrogen-induced polarization (PHIP) with the flexibility of polarization transfer via proton exchange. The new method can be used to polarize various molecules, including alcohols, water, lactate, and pyruvate. On average, only ≈3 mM of a hyperpolarized transfer agent was sufficient to significantly enhance the signal of ≈100 mM of target molecules via proton exchange. Thus, hydrogenative parahydrogen-induced hyperpolarization with proton exchange (PHIP-X) provides a new avenue for NMR applications beyond the limits imposed by thermal polarization.

Magnetic resonance (MR) is one of today's most versatile physical effects that is used in medical diagnostics[1] and chemical analysis[2] alike. Examples include everyday imaging in hospitals as well as the elucidation of the 3D structures of small molecules and proteins. In view of these powerful applications, which have been recognized by seven Nobel prizes, it may come as a surprise that no more than a few ppm of all spins in a sample effectively contribute to the MR signal. Indeed, the energy difference between the spin-up and spin-down states is orders of magnitude smaller than the thermal energy. According to the Boltzmann distribution, the two energy states are almost equally populated and only a tiny fraction, the polarization $P$, effectively contributes to the signal. To address this issue, ever stronger magnets are being developed, albeit at considerable cost (>1 M€). Still, even with the strongest magnets available today, more than 99.9 % of all spins remain invisible.

A different approach to this matter is being pursued via hyperpolarization. Here, various physical tricks are used to increase the polarization up to unity. The resulting signal gain has led to various breakthroughs in NMR and biomedical imaging and diagnostics[3–6]. Prominent examples include the detection of important reaction intermediates[4] and the early-stage detection of cancer[7], where hyperpolarized metabolites provide insights into metabolism, noninvasively, in vivo and in real-time.

The dominating technique for imaging hyperpolarized metabolites in vivo is dissolution dynamic nuclear polarization[8]. This method provides a high level of polarization (>50 %) of highly concentrated compounds (>100 mM); however, it requires complex and expensive hardware, including a superconducting magnet[9]. An alternative hyperpolarization approach takes advantage of the spin order of parahydrogen (pH$_2$). Here, polarization is achieved either by catalytically adding pH$_2$ or by reversible exchange[10]. Parahydrogen-induced hyperpolarization (PHIP)[3,11] has the advantage of providing a very fast hyperpolarization with inexpensive and simple equipment.

For a long time, however, all PHIP methods were limited to a few specific molecules[12,13]. The recent introduction of SABRE-Relay[14] has greatly expanded the range of molecules that can be hyperpolarized. In SABRE-Relay, a transfer agent with labile protons is hyperpolarized using SABRE. By exchanging these labile protons with labile protons of a target molecule hyperpolarization can be transferred to the target molecule. SABRE-Relay was analyzed in depth and optimized concerning field dependence, purity, and nature of solvent as well as the structure of the transfer agent and catalyst[15–17]. Still, the reported polarization level was far too low for biomedical imaging of metabolites[3,18] and the best performing transfer agents are toxic or corrosive. A closer inspection of the experimental evidence suggests that the low polarization of the transfer agent rather than the polarization transfer via proton exchange represents the bottleneck (in many cases, the polarization of the target molecule exceeded the polarization of the transfer agent[16]).

In this work, we present a new method (fig. 1) which combines the broad applicability of SABRE-Relay with the strong polarization achieved with hydrogenative PHIP. By direct hydrogenation of a transfer precursor, we were able to approach unity polarization on a transfer agent within seconds. The polarization is then transferred via proton exchange to various target molecules such as water, lactate, pyruvate and alcohols. This method, parahydrogen-induced polarization relayed via proton exchange (PHIP-X), provided much higher polarizations on a transfer agent than reported for SABRE-Relay[16]. On average, only ≈3 mM of the hyperpolarized transfer agent was able to significantly enhance the signal of ≈100 mM of target molecules via proton exchange.

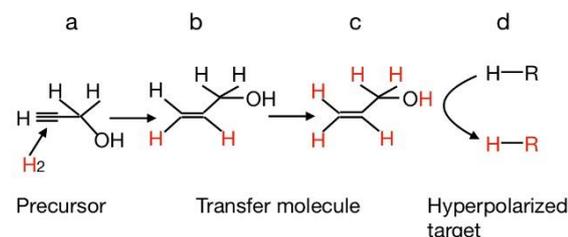

Figure 1: Schematic view of parahydrogen-induced polarization relayed via proton exchange (PHIP-X). Here, an unsaturated transfer agent (a) is hydrogenated with pH$_2$ (b). Net polarization is formed on all protons (c, red) and transferred to a target molecule (d) via proton exchange. The polarization achieved with the method is considerable: with 50 % enriched pH$_2$ more than 13 % of $^1$H-polarization was found on the transfer agent.

Here, we present the implementation of PHIP-X: First, pH$_2$ is catalytically added to an unsaturated precursor molecule at a low magnetic field (polarization field $B_p$). At this field, the hydrogens are strongly coupled and polarization is distributed among all coupled nuclear spins. Next, the sample is transferred to $B_0$ = 1 T, which is convenient for observation, as in adiabatic longitudinal transport after dissociation engenders nuclear alignment (ALTANDENA) experiments[19]. An interesting feature of the adiabatic field transfer is that it results in the net magnetization of the coupled spins[19]. Finally, the polarized labile protons of the transfer agent carry the net polarization to the target molecule via proton exchange. Thus, target molecules can be polarized without direct interaction with a catalyst or $pH_2$.

To identify a suitable system for PHIP-X, six different precursors of transfer agents (**1a** – **6a**, Fig. 2), three different solvents, and the catalyst $[Rh(dppb)(COD)]BF_4$ were evaluated (18 combinations). For these experiments, 7 mM of the catalyst and 85 mM of one precursor **1a** – **6a** (fig. 2) were dissolved in acetone-$d_6$, dichloromethane-$d_2$ (DCM-$d_2$), or DMSO-$d_6$ (if a substrate was not completely dissolved, the clean solution above the sediment was used instead). Hydrogen with 50 % of $pH_2$ fraction was bubbled at atmospheric pressure through 650 μl of each solution in a 5-mm NMR tube. Hydrogenation was performed in an external magnetic field of $B_p$ = 6 mT to allow spontaneous magnetization transfer (field dependency, see SI). After 60 seconds of $pH_2$ supply, the sample was manually placed into a bench-top NMR spectrometer ($B_0$ = 1 T, 43 MHz Spinsolve Carbon, Magritek), where a 90° pulse was applied shortly thereafter (additional details in SI).

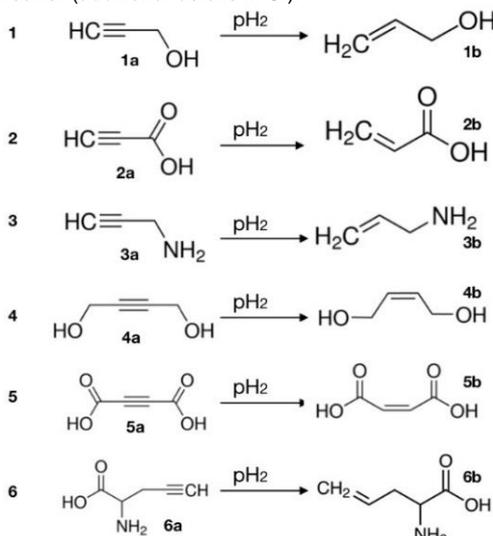

Figure 2. Structures of precursors and transfer agents used in this study. The following hydrogenation reactions were investigated: propargyl alcohol (**1a**) → allyl alcohol (**1b**), propiolic acid (**2a**) → acrylic acid (**2b**), propargyl amine (**3a**) → allyl amine (**3b**), butyne-1,4-diol (**4a**) → 2-butene-1,4-diol (**4b**), acetylenedicarboxylic acid (**5a**) → maleic acid (**5b**) and propargylglycine (**6a**) → allylglycine. (**6b**).

By far the strongest signal enhancement was found for propargyl alcohol **1a** in acetone-$d_6$ (fig. 3a). One of the peaks of proton H2 (proton no. 1 in tab. I) exhibited a 60,392-fold enhancement. Other precursors (**2a-6a**) resulted in a lower observable signal enhancement. Main reasons for the lower enhancements are a low reaction purity and bindings of functional groups the Rh+ atom of the catalyst. A detailed discussion of this is given in the supporting information (SI). For **1a** in acetone-$d_6$ we repeated the experiment with 20 s of $pH_2$ supply and obtained a ≈ 192,000-fold enhancement, owing to the fact that less than a third of the amount of **1b** contributes to the thermal spectrum taken for comparison. NMR in thermal equilibrium revealed that no more than 30 % of the precursor **1a** was hydrogenated (fig. 3b)**,** suggesting that the enhancement can be increased further. Therefore, later on we used only **1a** precursor to polarize transfer agent **1b** in acetone-$d_6$. It should be noted, too, that the aforementioned hydrogenation yield should be considered as an upper limit and polarization level as the lower limit because the hydrogenation continued after $pH_2$ supply was stopped due to residual $pH_2$ in solution (see SI).

For comparison with SABRE-Relay,[14] we integrated the entire multiplet of protons of transfer agent **1b** to quantify the polarization level. Strong polarization of more than 13 % was found on the H2 terminal proton of **1b** (tab. I). The important labile proton was polarized to more than 4 %. Note that this was achieved by using only 50% $pH_2$ enrichment; with 98 % $pH_2$ enrichment, we expect the polarization to be tripled[15] ($P_{HP}^*$ in table I). The highest polarizations for SABRE-Relay transfer agents found in the literature[16] are 0.49% for ammonia and 1.87% for $NH_2$- protons of benzyl-$d_7$-amine. Hence, the PHIP-X transfer agent allyl alcohol appears to be very promising.

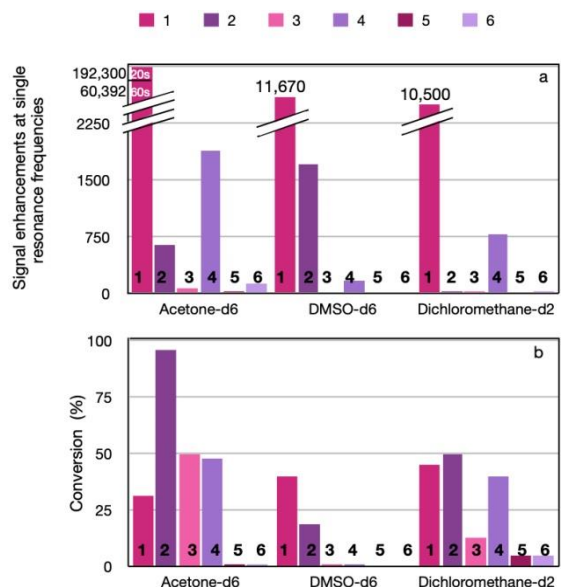

Figure 3. Enhancements of transfer agents **1b** – **6b** (a) and hydrogenation yield of precursors **1a** – **6a** (b) in three different solvents. Propargyl alcohol in acetone-D6 provided by far the strongest enhancements, 60,392-fold for 60 s and ≈ 192,000 fold for 20 s $pH_2$ supply at 1 T, but not the most efficient hydrogenation (about 30 %). Higher pressure is expected to increase the hydrogenation rate and thus produce larger amounts of transfer agent for the same time of hydrogenation. Note that enhancements refer to individual peaks within a multiplet (see SI for further information).

| | Measured and extrapolated polarization of the transfer agent. | | |
|---|---|---|---|
| | Proton | $P_{HP}$ [%] | $P_{HP}^*$ [%] |
| 1 | =CH– | >12,56 | >37,67 |
| 2 | H2C= | >13,31 | >39,93 |
| 3 | H2C= | >8,87 | >26,61 |
| 4 | –CH2– | >1,12 | >3,37 |
| 5 | OH | >4,18 | >12,35 |

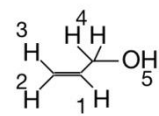

Allyl alcohol

Table I: Experimentally observed $^1H$ polarization $P_{HP}$ of individual protons (left) of allyl alcohol **1b** obtained in acetone-$d_6$ with 50% $pH_2$ enrichment and 20 s of $pH_2$ supply. (Right) The structure of **1b** with proton assignment. Reported values are lower limits because the hydrogenation fraction was measured 10 min after external $pH_2$ supply and the hydrogenation reaction was continuing due to dissolved $pH_2$. Increasing $pH_2$ enrichment to 98 % is expected to triple the polarization ($P_{HP}^*$).

To assess the efficacy of polarization transfer to a target molecule, we reduced the $pH_2$ supply duration to 20 s. This way, relaxation and further hydrogenation of **1b** to 1-

propanol was reduced. We exposed 85 mM of **1a** and 5 mM of catalyst in acetone-d6 for 20 s to 50 % $pH_2$ at $B_p = 6$ mT and transferred the sample to $B_0 = 1$ T (fig. 4a). Note that no extra target molecule was added, but that the precursor itself, **1a**, and residual water act as targets as they also possess exchanging labile protons. Consequently, all constituents, precursor (peaks 6 and 7), transfer molecule (peaks 1-5), and water (peak 9) were strongly polarized (fig. 4b, c). For quantification, 12,000 transients were acquired over 16 hours for the sample after hydrogenation. The polarizations of **1a** are presented in table I and the enhancement of protons 6 and 7 of **1b** was about 875-fold. The signal assignments of **1a**, **1b** and water are based on fig. S1 – S6 (SI).

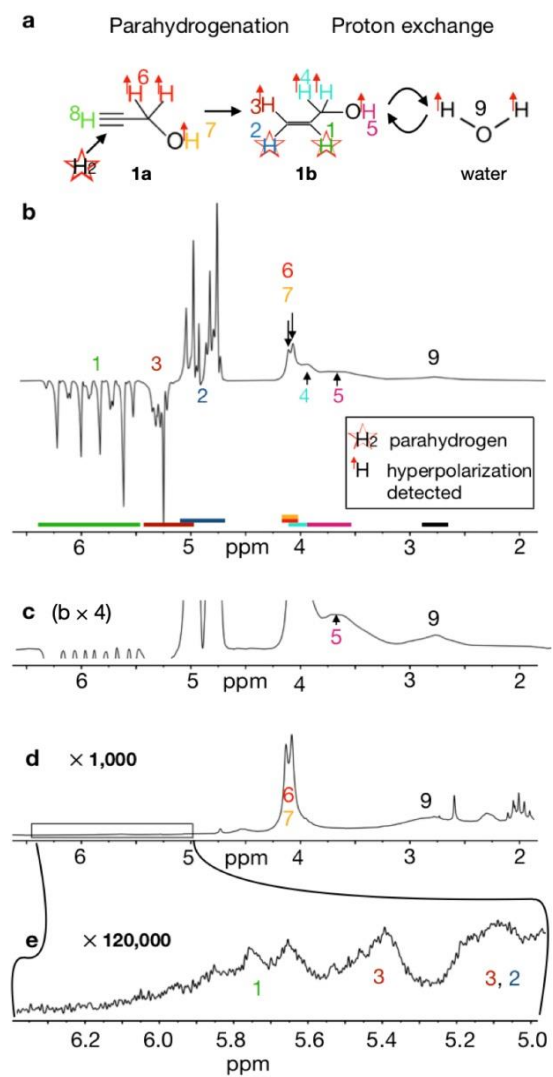

Figure 4. PHIP-X polarization of water and precursor **1a** as target molecules: polarization scheme (a), corresponding hyperpolarized (b, c) and thermal $^1$H NMR spectra (d, e, 12,000 transients, acquired 10 min after (b)). All constituents showed strong polarizations: P(**1b**, H5) > 4.18 %, P(**1b**, H6 and H7) = 0.4 % and P(H$_2$O) = 0.4 %. Note that the precursor itself featured a labile proton and acted as a target molecule. The very weak signals of the transfer agent (H1,2,3) in the thermal spectrum (d, e) indicate a low hydrogenation yield of less than 10 %. The spectrum in c is amplified by factor of 1,000 and the spectrum in e is amplified by a factor of 120,000. PHIP-X experimental parameters: 85 mM precursor **1a** and 5 mM catalyst were dissolved in acetone-d6 and exposed to 20 s of 50 % $pH_2$ at $B_p = 6$ mT before transfer to 1 T for signal acquisition after 90° flip angle excitation. Colored bars indicate the resonances of individual protons.

Next, we investigated the polarization transfer to ethanol. A solution containing 600 µl acetone-d6, 28 mM precursor **1a,** 28 mM ethanol, 48 mM (residual) water, and 5 mM [Rh(dppb)(COD)]BF$_4$ was polarized by PHIP-X as described above. Strongly enhanced signals of **1a, 1b**, ethanol, and water were detected (fig. 5b, c). A comparison of the data presented in fig. 4c and 5c proves that transfer of PHIP polarization to different molecules via proton exchange is feasible.

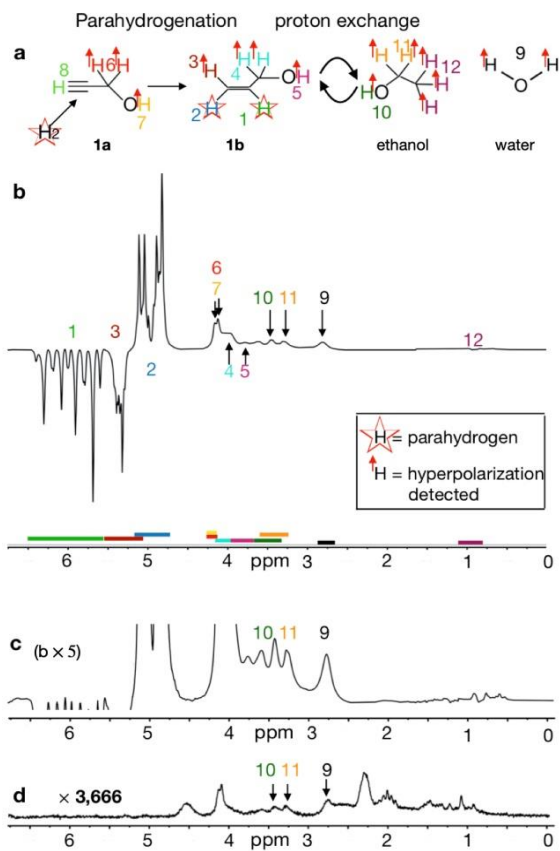

Figure 5. PHIP-X of ethanol and water with **1b** as a transfer agent: polarization scheme (a), corresponding hyperpolarized (b, c) and thermal $^1$H NMR spectra (d, 50 transients, acquired 10 min later). All constituents showed strong polarizations: P(ethanol, H11) = 0.42 %, P(H$_2$O) = 0.36 %. PHIP-X sample: 28 mM precursor, **1a**, 28 mM ethanol and 5 mM catalyst were dissolved in acetone-d6 and exposed to 20 s of 50% $pH_2$ at atmospheric pressure and $B_p = 6$ mT before transfer to 1 T and acquisition after 90° flip angle excitation.

For 28 mM of ethanol (fig. 6a) we found polarizations of 0.4 % for the OH-proton (H10), 0.42 % for the methylene group (H11), and 0.03 % for the methyl group (H12); 48 mM of water was polarized to 0.36 % (9); and approximately 25 mM of **1a** was polarized on average to 0.41 % for each proton of the CH$_2$OH group (H6,7). The polarization of the terminal HC≡ proton (H8) of **1a** could not be calculated because its signal overlaps with that of water; the shape of the corresponding NMR line indicates that water has a much higher polarization than H8 of **1a**.

On average, only ≈3 mM of the transfer agent **1b** was sufficient to significantly enhance the signal of ≈100 mM of target molecules. For SABRE-Relay[16] a 1:1 ratio of transfer agent to target molecule was found to be optimal for the polarization of target molecules. Thus, we expect to increase the polarization of target molecules further by increasing the hydrogenation pressure, temperature, and catalyst load.

In further experiments, lactic acid (LA) and pyruvic acid (PA) were polarized using PHIP-X with **1b** as a transfer agent. For 28 mM of LA a polarization of 0.07 % for proton H13 was obtained (fig. 6b, spectra in SI). For 120 mM of PA a polarization of 0.005 % was obtained for proton H14 (fig. 6c). Interestingly, we observed a much lower signal for **1b** than for ethanol. We hypothesize that the presence of LA and PA negatively influenced the hydrogenation of **1a** such that less polarized **1b** was observed. We suspect that LA and PA act as ligands for the Rh+ center of the catalyst and thus inhibit the catalytic cycle. To investigate this hypothesis, we repeated the experiment with PA but added PA after the parahydrogenation of **1a**, i.e., after stopping pH$_2$ supply. Compared to the process in which PA was added to the solution before hydrogenation, the signals from 1b were about ten times stronger. However, in this case the signal of the labile COOH proton overlaps strongly with other signals and one may estimate a polarization of PA (H14*) to 0.009 % with some unknown error (SI). An optimization of the hardware setup used for this experiment could decrease relaxation effects and further enhance the polarization. However, a higher concentration of **1b** resulted in a much broader signal of H14*, which overlaps with signals of **1b**.

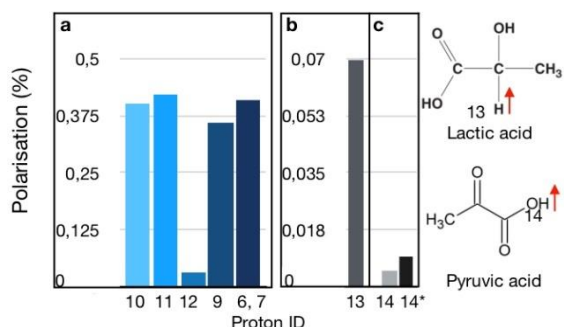

Figure 6: (a) Polarizations of 28 mM of ethanol, 84 mM of water, and 25 mM of **1a** achieved by PHIP-X (NMR spectra are shown in fig. 5). (b) Polarization of 29 mM of lactic acid (H13, right side) achieved by PHIP-X with **1b** as transfer agent. (c) Two different PHIP-X experiments were carried out for 120 mM of pyruvic acid, resulting in different polarizations for the COOH proton (H14, right side). If pyruvic acid was injected after (H14*) the parahydrogenation of **1a**, then larger signal amplifications were observed than when pyruvic acid was added before parahydrogenation of **1a**. The red arrows indicate the protons that were detected to be hyperpolarized via PHIP-X. The hyperpolarization of other protons could not be investigated due to overlapping signals.

Thus, the current experimental implementation of PHIP-X with **1a** in acetone-d6, 20 s hydrogenation at atmospheric pressure and manual transfer from $B_p$ = 6 mT to $B_0$ = 1 T, demonstrate the feasibility of PHIP-X as a new approach to broaden the applicability of hydrogenating PHIP. We foresee that further optimization for individual tracers such as PA and LA will greatly improve the polarization yield. Apparent routes to optimize the polarization yield also include (i) a variation (deuteration) of the transfer agents and catalyst, (ii) optimization of polarization transfer conditions (e.g., adjustment of an external magnetic field, $B_p$ (SI), and magnetic field trajectory to $B_0$), (iii) purity and composition of the solvent system, (iv) higher pH$_2$ enrichment, and (v) higher hydrogenation pressure. For now, we can extrapolate the effect of the latter two. An increase from 50 % to 98 % pH$_2$ fraction usually results in a threefold increase in signal amplification.[15] The hydrogenation is affected, for example, by the choice of catalyst, hydrogenation pressure, and temperature. Higher values of pressure and temperature usually result in faster hydrogenation[19], smaller relaxation effects (shorter hydrogenation time), and higher polarization. Thus, assuming a threefold increase by using 98% pH$_2$ and another threefold increase by using 3 bar instead of 1 bar (which should hold up to a certain level but remains speculative), one order of magnitude higher polarization may be achieved, which would result in polarizations of target molecules in the range of a few percentage points.

To summarize, PHIP-X is a new and very general method for hyperpolarization of diverse molecules, including alcohols, water, and biomolecules such as pyruvate and lactate. Combining the wide applicability of SABRE-Relay and the strong signal amplification of PHIP, PHIP-X opens new avenues to hyperpolarize principally every molecule with labile protons within seconds, at room temperature, in the liquid state, and without chemical changes of the target molecule.

**Acknowledgements**

We acknowledge support by the Emmy Noether Program "metabolic and molecular MR" (HO 4604/2-2), the research training circle "materials for brain" (GRK 2154/1-2019), DFG-RFBR grant (HO 4604/3-1, No 19-53-12013), Cluster of Excellence "precision medicine in inflammation" (PMI 1267). Kiel University and the Medical Faculty are acknowledged for supporting the Molecular Imaging North Competence Center (MOIN CC, MOIN 4604/3). MOIN CC was founded by a grant from the European Regional Development Fund (ERDF) and the Zukunftsprogramm Wirtschaft of Schleswig-Holstein (Project no. 122-09-053) The Russian team thanks the Russian Foundation for Basic Research (Grant 19-53-12013) for financial support.

# Supplementary Information for

**Parahydrogen-induced polarization relayed via proton exchange.**

Kolja Them[1,6], Frowin Ellermann[1], Andrey N. Pravdivtsev[1], Oleg G. Salnikov[2,3,4], Ivan V. Skovpin[2,3], Igor V. Koptyug[2,3], Rainer Herges[5] and Jan-Bernd Hövener[1,7]

[1]Section Biomedical Imaging, Molecular Imaging North Competence Center (MOIN CC), Department of Radiology and Neuroradiology, University Medical Center Schleswig-Holstein and Kiel University, Am Botanischen Garten 14, 24118 Kiel, Germany

[2]International Tomography Center, SB RAS, 3A Institutskaya st., Novosibirsk 630090, Russia

[3]Novosibirsk State University, 2 Pirogova st., Novosibirsk 630090, Russia

[4]Boreskov Institute of Catalysis SB RAS, 4 Acad. Lavrentiev pr., Novosibirsk 630090, Russia

[5]Otto-Diels-Institute for Organic Chemistry, Christian-Albrechts-University, 24118 Kiel, Germany

[6]kolja.them@rad.uni-kiel.de, [7]jan.hoevener@rad.uni-kiel.de

Content:

1. Materials and Methods.

2. $^1$H NMR spectra for signal assignments for propargyl alcohol and allyl alcohol in acetone-$d_6$.

3. Reaction purity for the catalytic hydrogenation of propargyl alcohol in acetone-$d_6$.

4. Signal enhancement for transfer agents 1b – 6b and hydrogenation efficiency for precursors 1a to 6a in acetone-$d_6$, dichloromethane-$d_2$ and dimethyl sulfoxide-$d_6$.

5. Lactic acid hyperpolarized by PHIP-X.

6. Pyruvic acid hyperpolarized by PHIP-X.

7. Dependence of the NMR signal enhancement on the polarization field in PHIP-X.

1. Materials and Methods

All chemicals except solvents were purchased from Sigma Aldrich and used without further purification. Solvents were purchased from Deutero and used without further purification. All NMR measurements were carried out using a 5mm NMR tube and a 1-T bench-top NMR spectrometer (Spinsolve Carbon, Magritek). $pH_2$ with about 50 % enrichment was produced using a home-built $pH_2$ generator with liquid nitrogen as cooling source. All detailed information on concentrations of chemicals, bubbling time, and polarization fields are provided at the corresponding places in the main manuscript or SI. Detailed information on the general PHIP-X method is presented in the main manuscript.

2. ¹H NMR spectra for signal assignments for propargyl alcohol and allyl alcohol in acetone-d6.

Here we present the thermal ¹H NMR spectra that are necessary for peak assignments of propargyl alcohol (precursor of transfer agent), allyl alcohol (transfer agent), and water in acetone-d6. The signal of a small amount of water (naturally contained in the solvent acetone-d6) is located at 2.85 ppm and the five peaks of residual protons of acetone are centered at 2.05 ppm. For the investigation of the reaction purity, it is important to analyze the region from 4.8 ppm to 6.4 ppm, which clearly shows that parahydrogenation of propargyl alcohol leads to hyperpolarized allyl alcohol with high purity. Deuterated water is added to some solutions in order to identify nonlabile protons. One important ¹H NMR spectrum is the superposition of propargyl alcohol and allyl alcohol (both with 10 μl D$_2$O) in acetone-d6. Based on this spectrum, most of the NMR signals of the transfer agent and the precursor in PHIP-X experiments can be identified. The important labile proton can be identified by comparing 5 μl allyl alcohol in acetone-d6 and 5 μl allyl alcohol with 10 μl D$_2$O in acetone-d6.

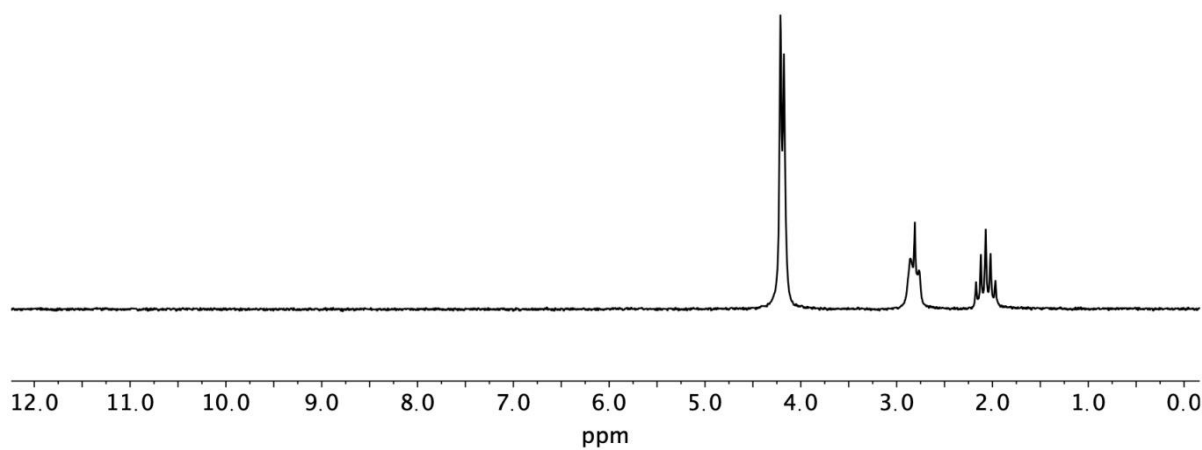

Fig. S1: Thermal ¹H NMR spectrum of 5 μl propargyl alcohol in 500 μl acetone-d6. A comparison with S2 clarifies the signals.

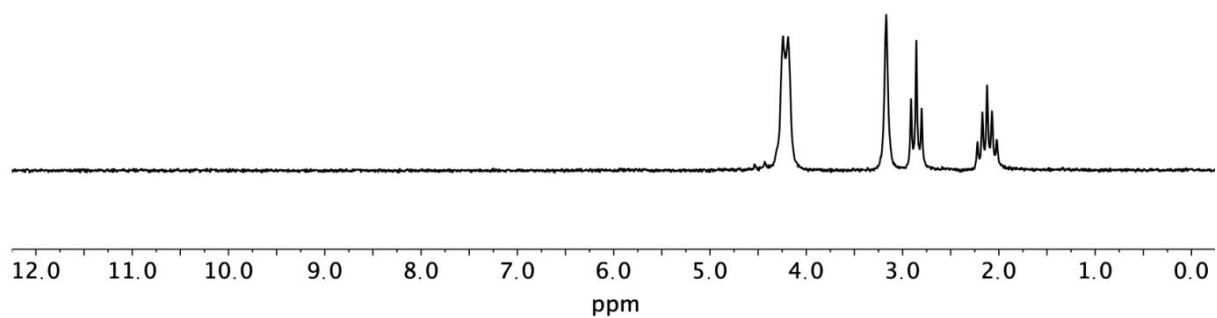

Fig. S2: Thermal ¹H NMR spectrum of 5 μl propargyl alcohol with 10 μl D$_2$O in 500 μl acetone-d6. Comparing this figure with S1, it can be seen that the signal of the labile OH proton strongly overlaps with -CH$_2$- protons. The signal of the HC≡ proton is located at 2.77 ppm. Due to the addition of 10 μl D$_2$O the signal of water is shifted from 2.85 ppm to 3.10 ppm.

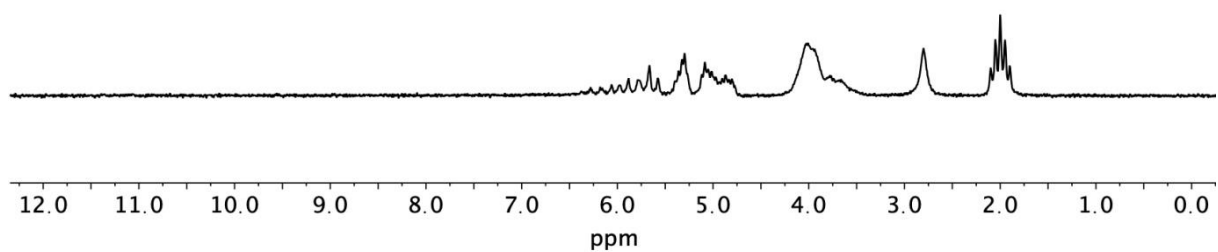

Fig. S3: Thermal ¹H NMR spectrum of 5 µl allyl alcohol in 500 µl acetone-d6. Compared to the case where OH of allyl alcohol is replaced by the OD one estimates the OH signal between 3.5 and 4.0 ppm.

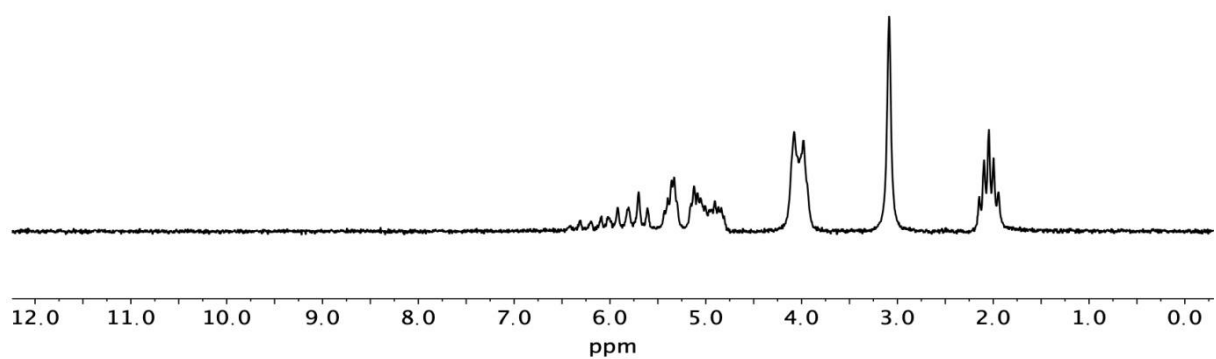

Fig. S4: Thermal ¹H NMR spectrum of 5 µl allyl alcohol with 10 µl D₂O in 500 µl acetone-d6. Due to the addition of 10 µl D₂O, the signal of water is shifted from 2.85 ppm to 3.10 ppm while the labile OH proton of allyl alcohol vanishes due to deuteration.

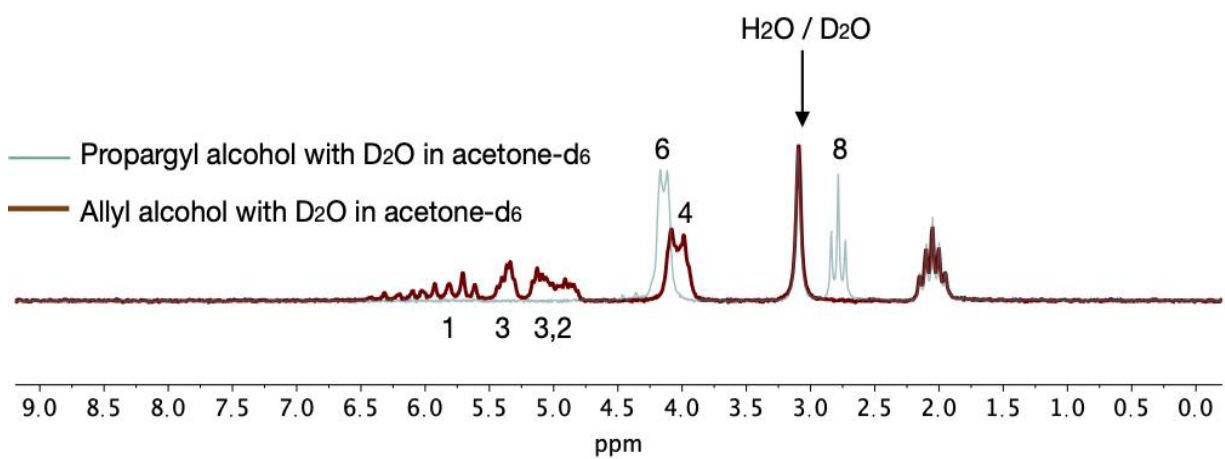

Fig. S5: Superposition of two thermal ¹H NMR spectra: 5 µl propargyl alcohol with 10 µl D₂O in 500 µl acetone-d6 (fig S2) and 5 µl allyl alcohol with 10 µl D₂O in 500 µl acetone-d6 (fig. S4). The numbers located at the NMR signals correspond to the numbering scheme of the main manuscript. The region from 4.75 - 6.4 ppm contains only signals from ally alcohol. It is important to compare the -CH₂- signals of propargyl alcohol (6) and allyl alcohol (4). The signal associated with propargyl alcohol is slightly more deshielded. This fact is important for the analysis of PHIP-X spectra because both 4 and 6 are hyperpolarized. Signal assignments for 1, 2, 3 and 4 can be found in a public database, e.g., www.chemicalbook.com.

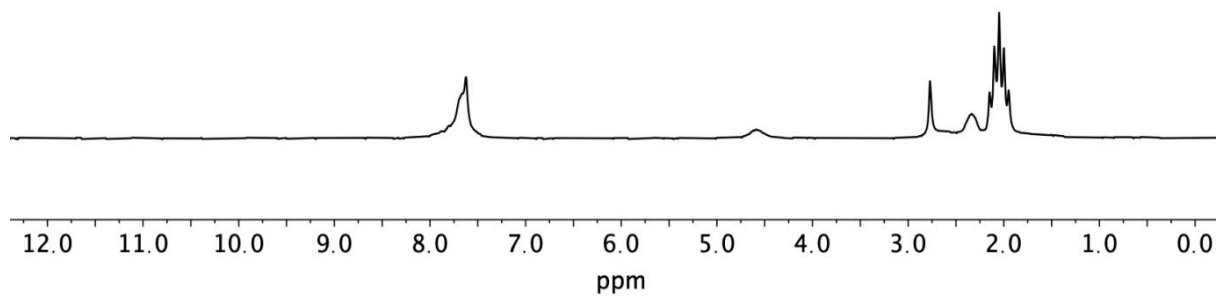

Fig. S6: Thermal ¹H NMR spectrum of 5 mM of catalyst [Rh(dppb)(COD)]BF₄ in acetone-d₆. The catalyst has nearly no overlapping signals with propargyl alcohol, allyl alcohol, and water.

3. Reaction purity

In order to investigate the reaction purity for the catalytic hydrogenation of propargyl alcohol to allyl alcohol with the conditions present in PHIP-X experiments, we bubbled hydrogen at atmospheric pressure for 60 seconds through a solution of 5 mM μl propargyl alcohol with 5 mM catalyst [Rh(dppb)(COD)]BF₄ in acetone-d₆. A comparison with a thermal ¹H NMR spectrum of allyl alcohol in acetone-d₆ was used to investigate the reaction purity. The result showed a purity of about 72 % in the thermal spectrum. Due to overlapping signals there might be an error, which is difficult to estimate. The full spectrum from 0 - 12 ppm (fig. S7 a) together with a zoom (fig. 7b) to 4.8 - 7.0 ppm show that mainly allyl alcohol is produced in PHIP-X conditions. A main impurity is derived from the production of 1-propanol from a further hydrogenation of allyl alcohol. The production of 1-propanol depends on the duration of parahydrogen supply. After 120 s hyperpolarized 1-propanol is visible in PHIP-X spectra. However, in the hyperpolarized PHIP-X spectra shown in this document, no signals of side products were found which affected the data analysis.

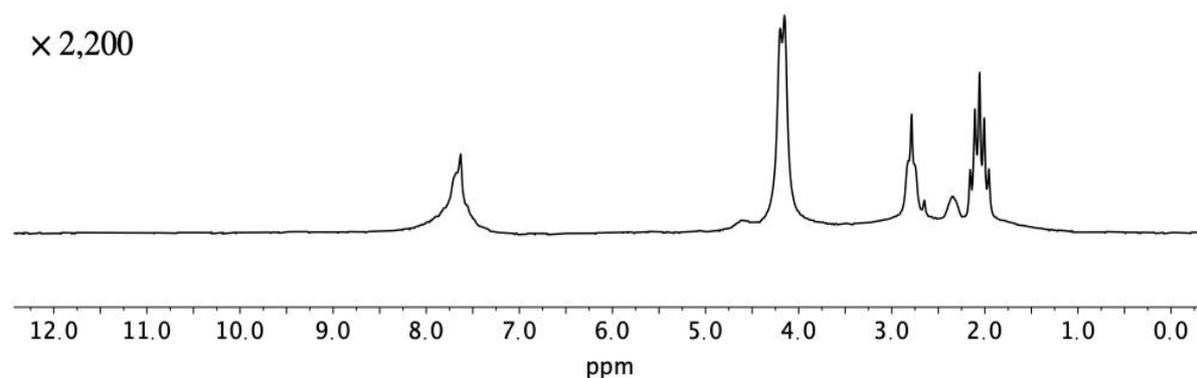

S7a: Thermal ¹H NMR spectrum of 85 mM propargyl alcohol in acetone-d₆ with 5.5 mM [Rh(dppb)(COD)]BF₄. Compared to the corresponding hyperpolarized spectrum, this spectrum is enlarged by a factor of 2,200.

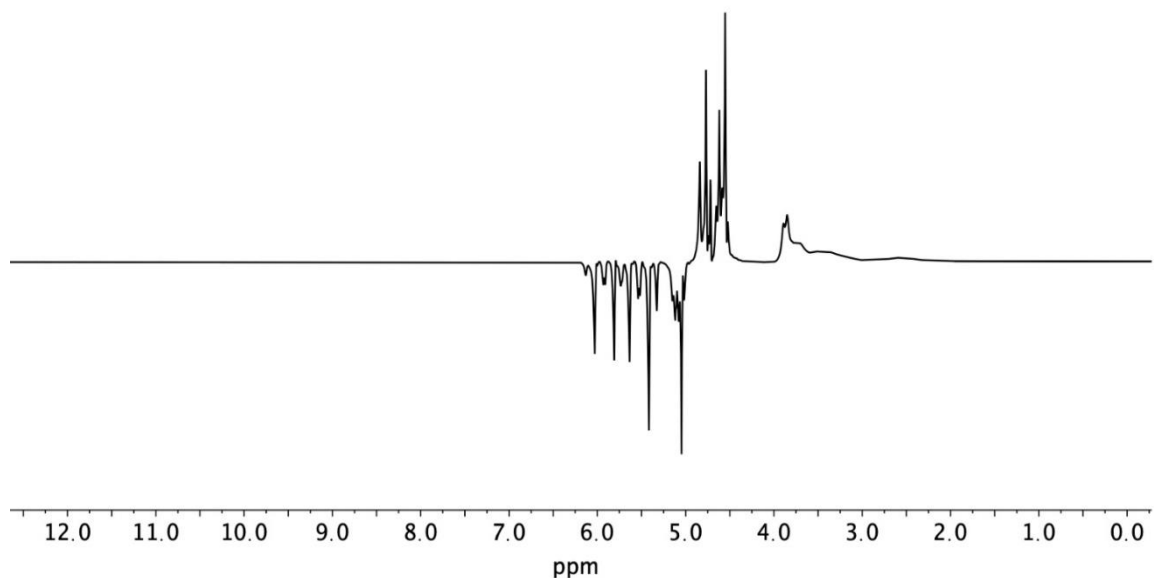

S7b: Hyperpolarized $^1$H NMR spectrum of hydrogenated propargyl alcohol in acetone-$d_6$.

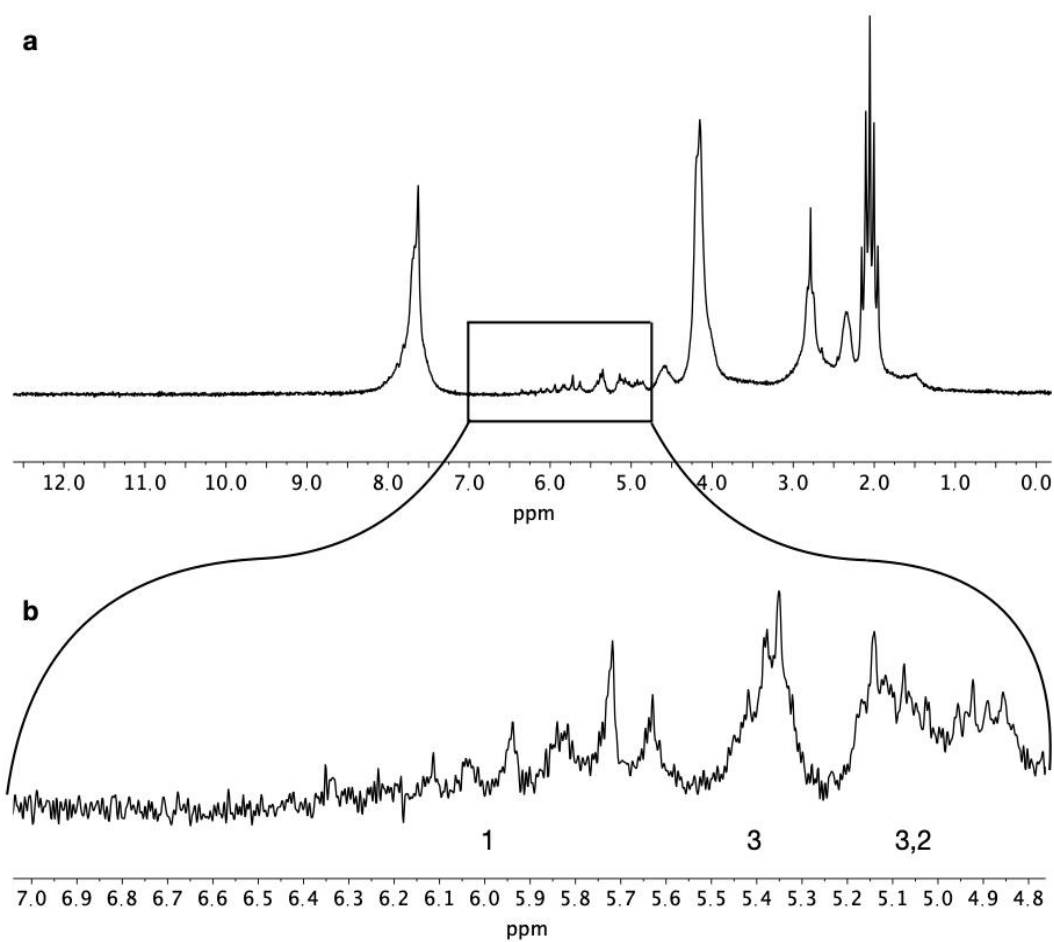

Fig. S7c: (a) Thermal $^1$H NMR spectrum acquired 10 minutes after parahydrogenation of propargyl alcohol to allyl alcohol with conditions as in PHIP-X experiments. (b) The important region from 4.8 ppm to 7.0 ppm is enlarged. A comparison with allyl alcohol in acetone-$d_6$ shows that the reaction purity for this hydrogenation reaction is at least 72 %. Compared to the corresponding hyperpolarized spectrum in S7b, the spectrum of S7c is enlagred by a factor of 4,600.

4. Signal enhancements and hydrogenation efficiency for transfer agents.

In order to identify a suitable system for PHIP-X, six different precursors (1a – 6a, Fig. 2 of the main manuscript), three different solvents, and the catalyst [Rh(dppb)(COD)]BF4 were evaluated (18 combinations). For these experiments, 7 mM of catalyst and 85 mM precursor 1a to 6a (fig. 2) were dissolved in acetone-d6, dichloromethane-$d_2$ (DCM-$d_2$), or DMSO-$d_6$. If a substrate was not completely dissolved, the clean solution above the sediment was used instead. Hydrogen with 50 % p$H_2$ enrichment was bubbled at atmospheric pressure through 650 µl of each solution in a 5-mm NMR tube. Bubbling was performed in an external magnetic field of $B_p$ = 6 mT in order to allow spontaneous magnetization transfer. After 60 seconds of p$H_2$ supply, the sample was manually transferred into a $B_0$ = 1 T bench-top NMR spectrometer (Spinsolve Carbon, Magritek), where a 90° pulse was applied shortly thereafter. The following signal enhancements, shown in figure 3a of the main manuscript, were found:

Allyl alcohol 1b in acetone-$d_6$ exhibited a 192,000-fold enhancement. In DMSO-d6 1b showed a maximum signal enhancement of 11,670 and in DCM-$d_2$ of 10,500. Hydrogenation of propiolic acid 2a to acrylic acid 2b provided enhancements of 640 in acetone-$d_6$, 1700 in DMSO-d6, and 26 in DCM-$d_2$. Hydrogenation of 2-butyne-1,4-diol 4a to 2-butene-1,4-diol 4b provided enhancements of 1,887 in acetone-d6, 175 in DMSO-$d_6$, and 775 in DCM-$d_2$. Hydrogenation of propargylamine 3a to allyl amine 3b provided 76 in acetone-$d_6$, 0 in DMSO-$d_6$ and 23 in DCM-$d_2$. A reason for the low signal enhancements is due to binding of the amine group to the Rh+ of the catalyst. Hydrogenation of acetylenedicarboxylic acid 5a to maleic acid 5b provided enhancements of 30 in acetone-$d_6$, 0 in DMSO-$d_6$, and 18 in DCM-$d_2$. A reason for this very low enhancement is very poor solubility in the organic solvents used. Hydrogenation of propargylglycine 6a to allylglycine 6b provided enhancements of 130 in acetone-d6, 0 in DMSO-$d_6$, and 24 in DCM-$d_2$. A reason for the low enhancements may be a binding of the amine group to the Rh+ of the catalyst.

**Reasons for the choice of 1a for the generation of a transfer agent:**

The two major reasons for the choice of 1a are the most strongest signal enhancement and a relatively high reaction purity which holds for a sufficient long time of p$H_2$ supply. The consumption of about 30% after 60 seconds of p$H_2$ supply to a solution of 85 mM means that there is a strong potential for an optimization of the method by increasing the hydrogenation pressure, temperature and catalyst load. This would increase the speed of the hydrogenation reaction which allows to produce a larger amount of the transfer agent in a shorter time. Hence, relaxation effects will be reduced and there is more polarization in total (molar polarization), which can be distributed to target molecules.

From table I of the main manuscript one may think that 1a in DCM-$d_2$ and DMSO-$d_6$ are promising systems as well. However, looking at the corresponding spectra it can be seen that the reaction purity is much more worse: After 60 seconds of p$H_2$ supply only very small amounts of 1b are generated although a lot of 1a is consumed. It seems that in DCM-$d_2$ and DMSO-$d_6$ 1b is much faster hydrogenated to 1-propanol and/or more side products are produced compared to the case of acetone-$d_6$. For example, if the hydrogenation of 1a in acetone-$d_6$ is stopped after 20 s an almost three-fold stronger signal enhancement is obtained compared to the case of 60 s. This is because in 20 s roughly a third of the amount of 1b is produced compared to the case of 60 s. However, we did not observed this effect for 1a in DCM-$d_2$ and DMSO-$d_6$, because of the low reaction purity in these solvents. Furthermore, we did not observed large amounts of strongly polarized 1-propanol for DCM-$d_2$ and DMSO-$d_6$. Therefore, DCM-$d_2$ and DMSO-$d_6$ seem to be unsuitable to produce large amounts of a hyperpolarized transfer agents.

The parahydrogenation of precursors containing an NH2-group (3a and 6a) suffers from bindings between the free electron pair of the nitrogen atom to the positively charged metal ion of the catalyst. Such a binding event prevents that the tripple bond comes close to the metal ion. Furthermore, a low reaction purity was observed. Precursor 4a provides the highly symmetric transfer agent 4b. The symmetry is a draw back for observable signals, because p$H_2$ spin order is not broken. Precursor 5a suffers from very low solubility in the used solvents.

There are two major reasons for choosing **1a** to generate a transfer agent: Firstly, it provides the strongest signal enhancement and a relatively high reaction purity, which holds for a sufficiently long duration of pH2 supply. The consumption of about 30% after 60 seconds of pH2 supply to a solution of 85 mM means that there is a strong potential for optimizing the method by increasing the hydrogenation pressure, temperature, and catalyst load. This would increase the speed of the hydrogenation reaction, which can then produce a larger amount of the transfer agent in a shorter time. Hence, relaxation effects will be reduced and there is more polarization in total (molar polarization), which can be distributed to target molecules.

From table I of the main manuscript one might assume that **1a** in DCM-$d_2$ and DMSO-$d_6$ are promising systems as well. However, looking at the corresponding spectra the reaction purity is clearly much worse: After 60 seconds of pH2 supply, only very small amounts of **1b** are generated although a considerable amount of **1a** is consumed. It seems that in DCM-$d_2$ and DMSO-$d_6$ **1b** is hydrogenated to 1-propanol much faster and/or more side products are produced than in acetone-$d_6$. For example, if the hydrogenation of **1a** in acetone-$d_6$ is stopped after 20 s, an almost three-fold stronger signal enhancement is obtained than after 60 s. This is because in 20 s roughly a third of the amount of **1b** is produced compared to the case of 60 s. However, we did not observe this effect for **1a** in DCM-$d_2$ and DMSO-d6 owing to the low reaction purity in these solvents. Furthermore, we did not observe large amounts of strongly polarized 1-propanol for DCM-$d_2$ and DMSO-$d_6$. Therefore, DCM-$d_2$ and DMSO-$d_6$ seem to be unsuitable to produce large amounts of a hyperpolarized transfer agent.

The parahydrogenation of precursors containing an $NH_2$-group (**3a** and **6a**) suffers from bindings between the free electron pair of the nitrogen atom to the positively charged metal ion of the catalyst. Such a binding event prevents the triple bond from getting close to the metal ion. Furthermore, a low reaction purity was observed. Precursor **4a** provides the highly symmetric transfer agent **4b**. The symmetry is a drawback for observable signals because $pH_2$ spin order is not broken. Precursor **5a** suffers from very low solubility in the solvents that were used.

For each of the 18 experiments we now show three $^1H$ NMR spectra. First, a thermal $^1H$ NMR spectrum is always recorded shortly before the $pH_2$ supply (500 scans). This is followed by a $^1H$ NMR spectra recorded immediately after the parahydrogenation. Then, a thermal $^1H$ NMR spectrum is recorded 10 minutes after the $pH_2$ supply (500 scans). The spectra for propargyl alcohol are shown in the main manuscript (figure 4). Thermal and hyperpolarized $^1H$ NMR spectra of the precursors and transfer agents are shown below:

**Parahydrogenation in acetone-$d_6$:**

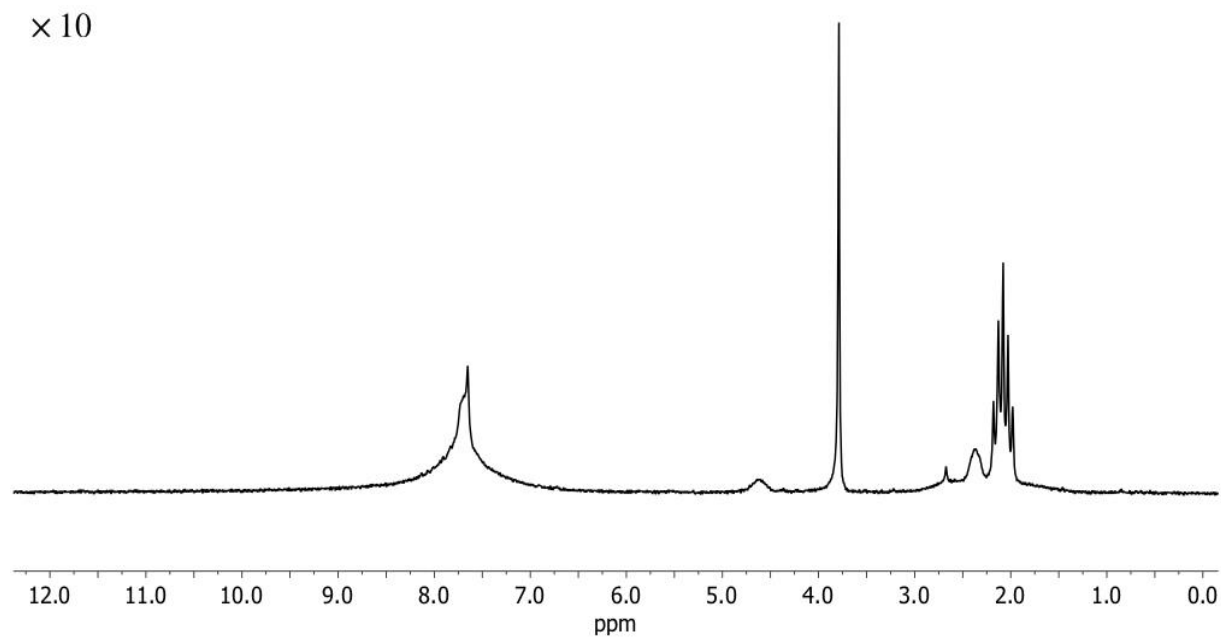

S8: Thermal $^1H$ NMR spectrum of 85 mM propiolic acid in acetone-$d_6$ with 5.5 mM [Rh(dppb)(COD)]BF4. Compared to the corresponding hyperpolarized spectrum, this spectrum is enlarged by a factor of 10.

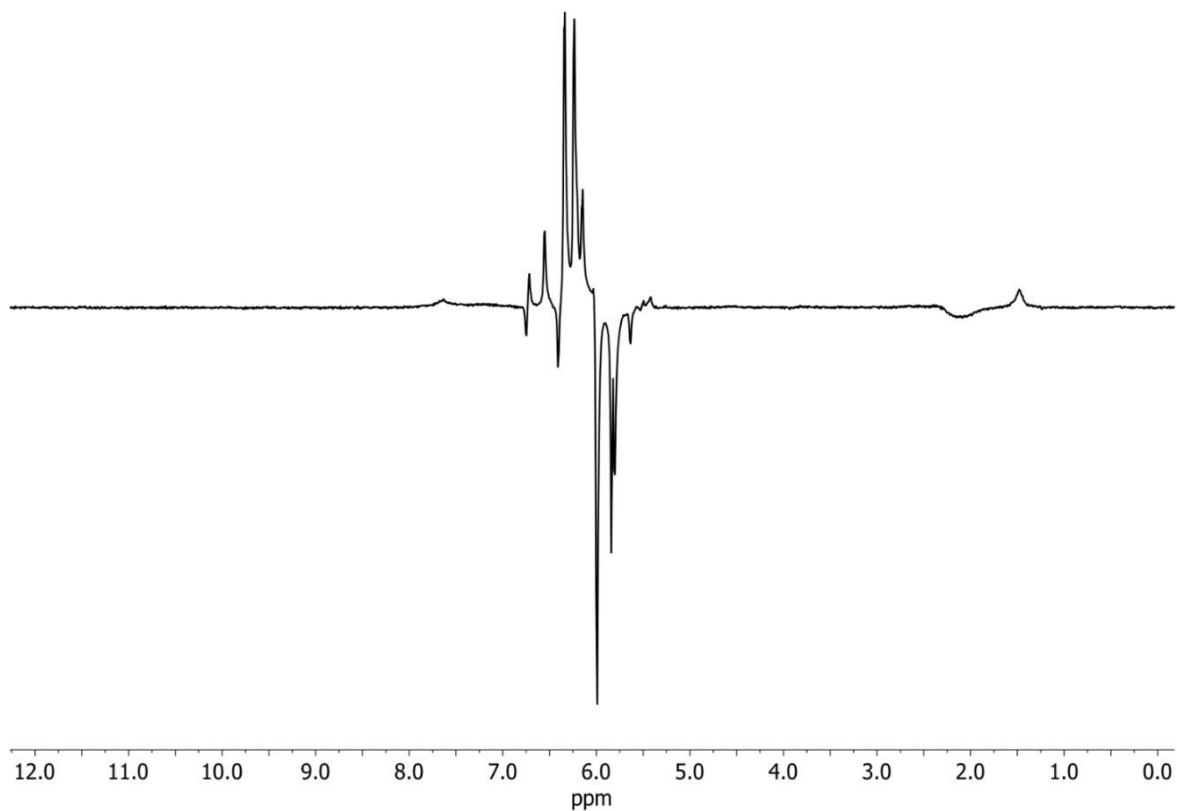

S9: Hyperpolarized $^1$H NMR spectrum of hydrogenated propiolic acid in acetone-$d_6$.

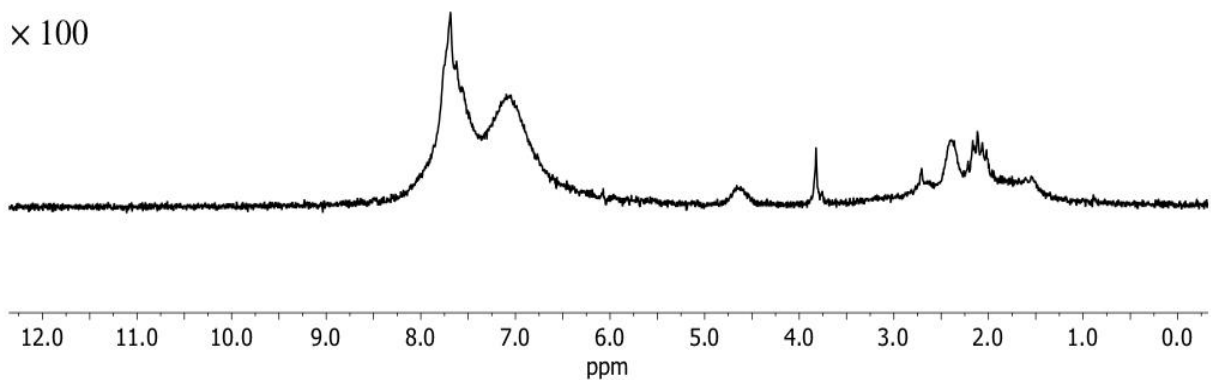

S10: Thermal $^1$H NMR spectrum recorded 10 minutes after the acquisition of S9. Compared to the corresponding hyperpolarized spectrum, this spectrum is enlarged by a factor of 100.

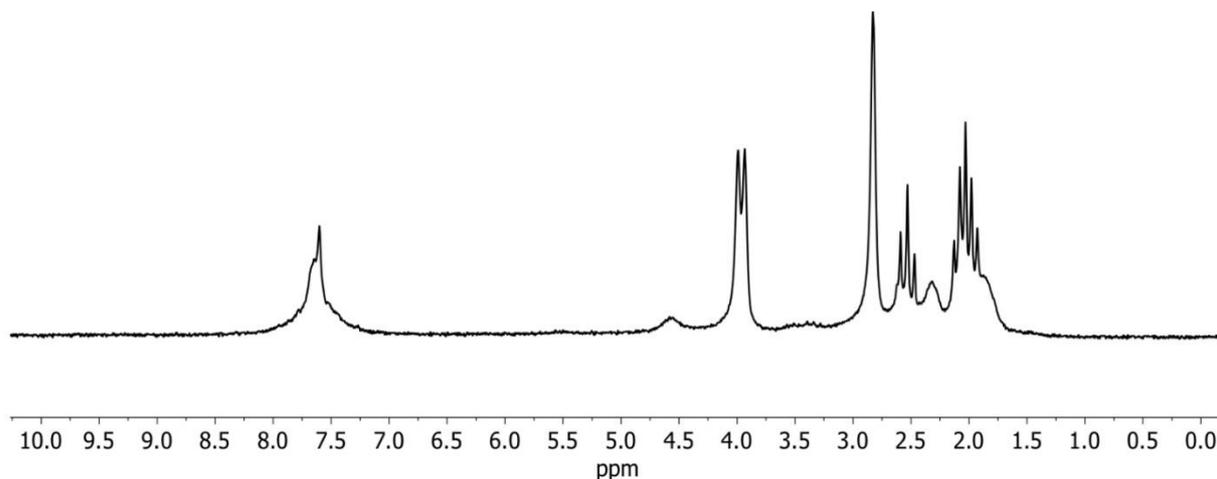

S11: Thermal ¹H NMR spectrum of 85 mM propargyl amine in acetone-d$_6$ with 5.5 mM [Rh(dppb)(COD)]BF4. Compared to the corresponding hyperpolarized spectrum, this spectrum is not enlarged.

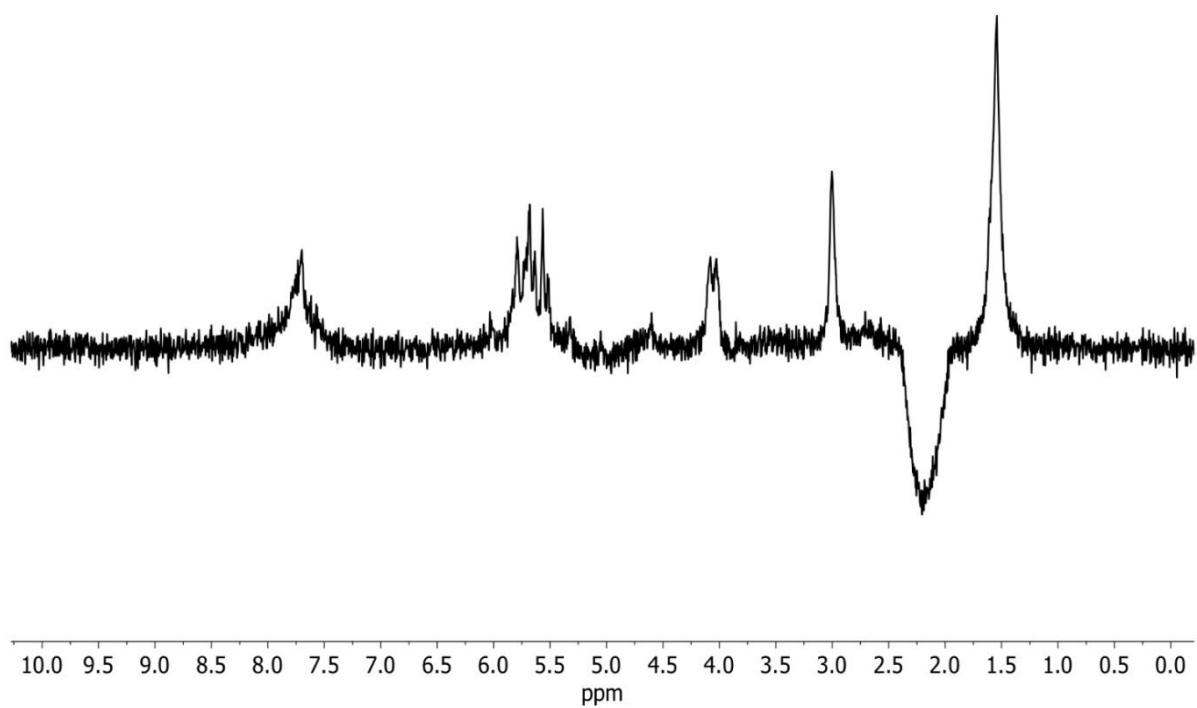

S12: Hyperpolarized ¹H NMR spectrum of hydrogenated propargyl amine in acetone-d$_6$.

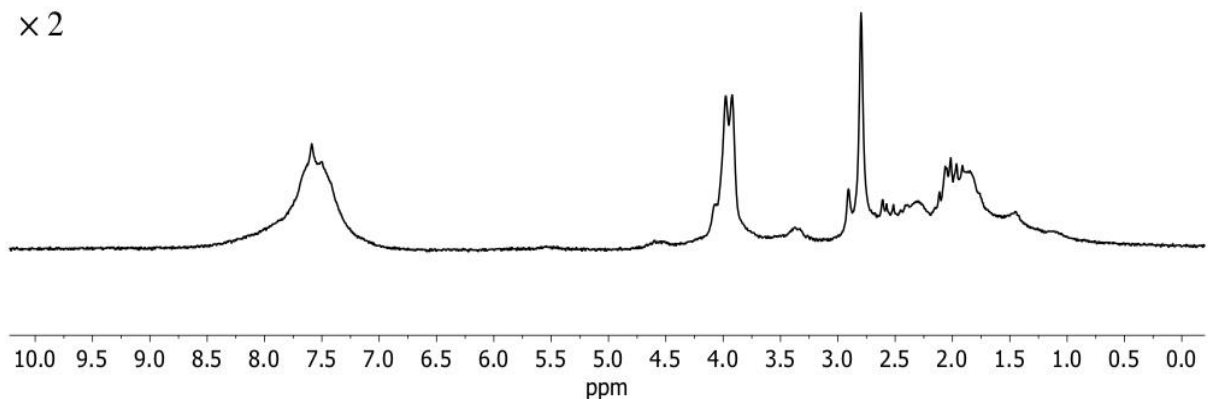

S13: Thermal ¹H NMR spectrum recorded 10 minutes after the acquisition of S12. Compared to the corresponding hyperpolarized spectrum, this spectrum is enlarged by a factor of 2.

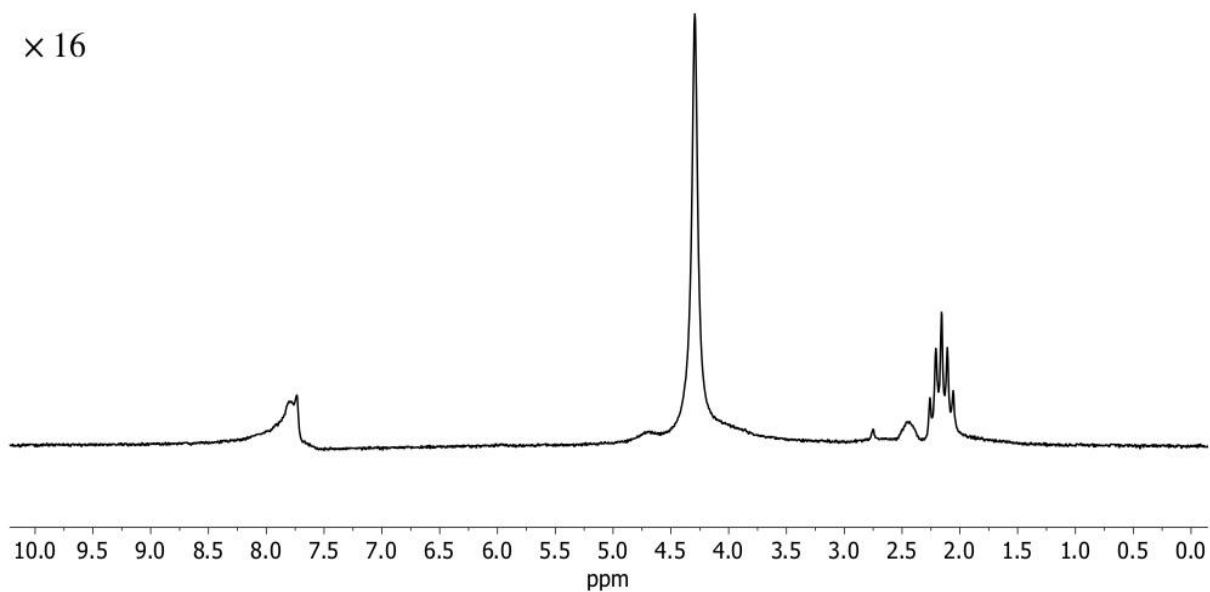

S14: Thermal ¹H NMR spectrum of 85 mM butyne-1,4-diol in acetone-$d_6$ with 5.5 mM [Rh(dppb)(COD)]BF4. Compared to the corresponding hyperpolarized spectrum, this spectrum enlarged by a factor of 16.

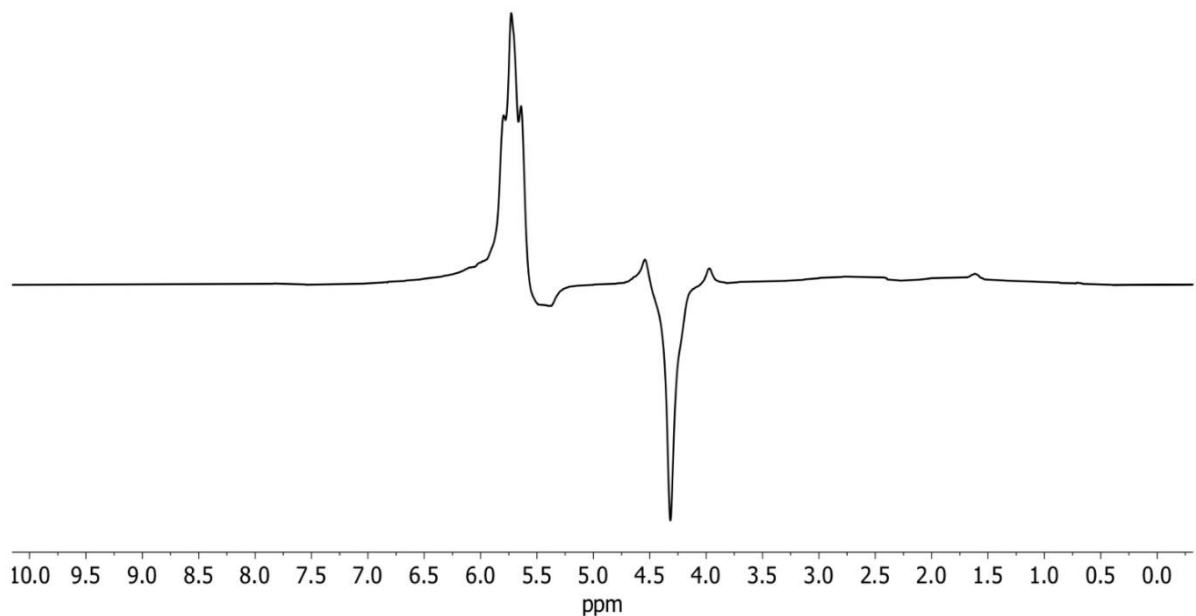

S15: Hyperpolarized $^1$H NMR spectrum of hydrogenated butyne-1,4-diol in acetone-$d_6$.

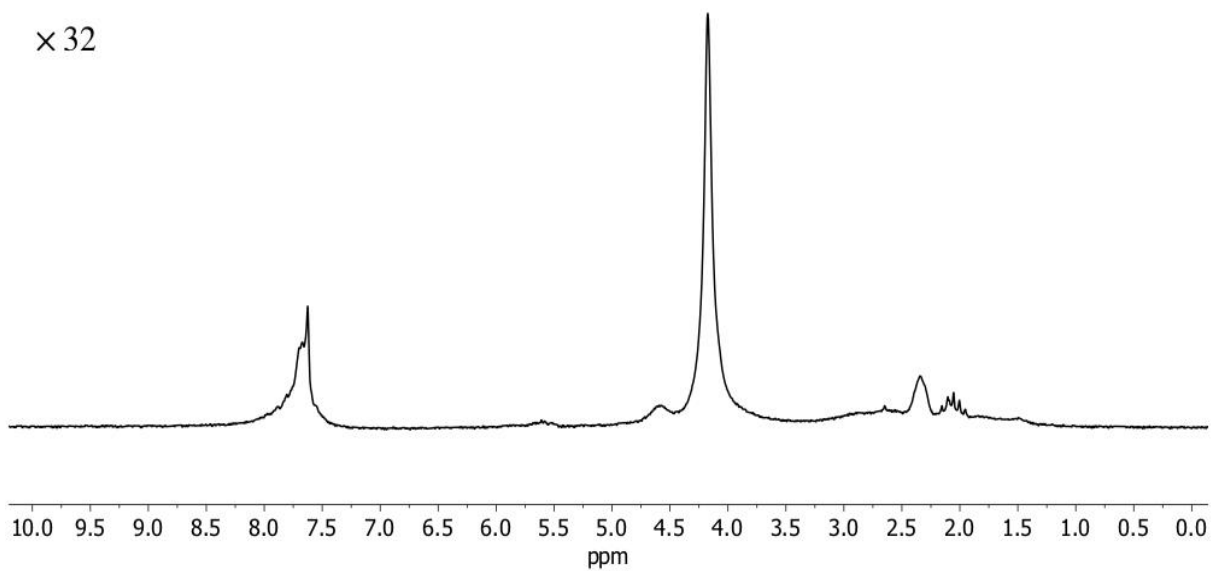

S16: Thermal $^1$H NMR spectrum recorded 10 minutes after the acquisition of S15. Compared to the corresponding hyperpolarized spectrum, this spectrum is enlarged by a factor of 32.

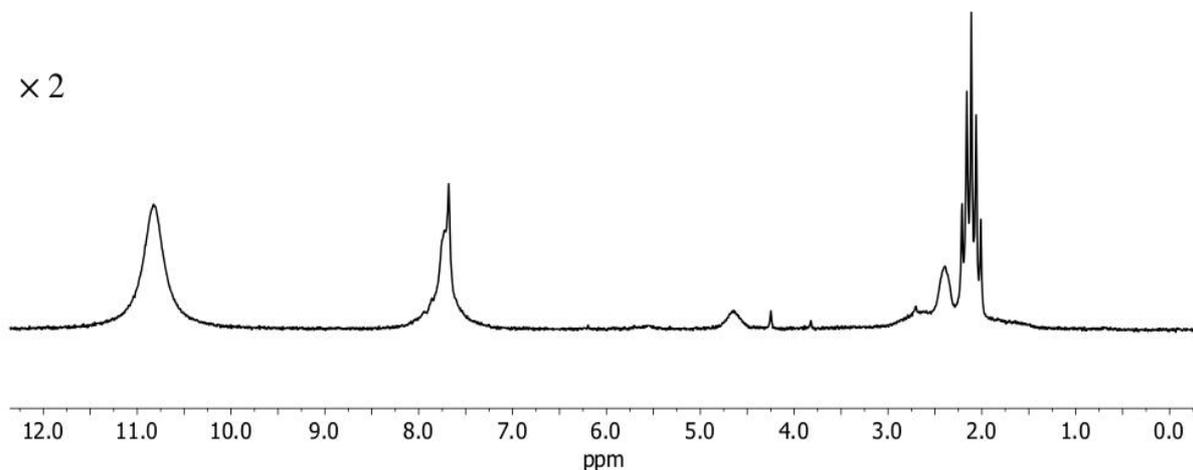

S17: Thermal ¹H NMR spectrum of 85 mM acetylenedicarboxylic acid in acetone-d₆ with 5.5 mM [Rh(dppb)(COD)]BF4. Compared to the corresponding hyperpolarized spectrum, this spectrum enlarged by a factor of 2.

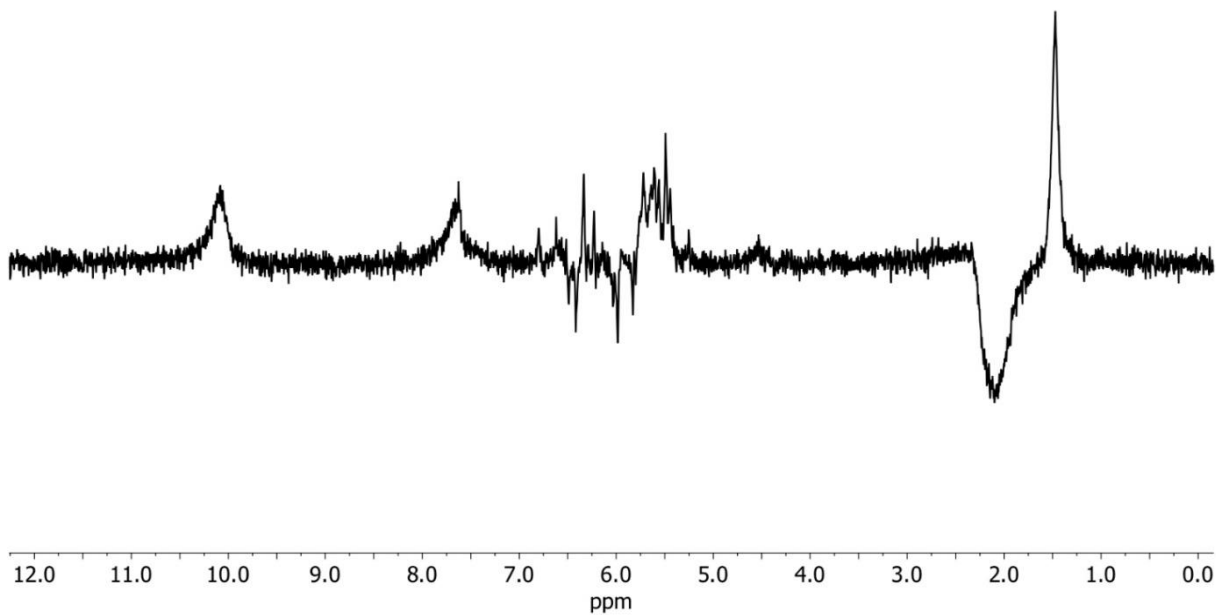

S18: Hyperpolarized ¹H NMR spectrum of hydrogenated acetylenedicarboxylic acid in acetone-d₆.

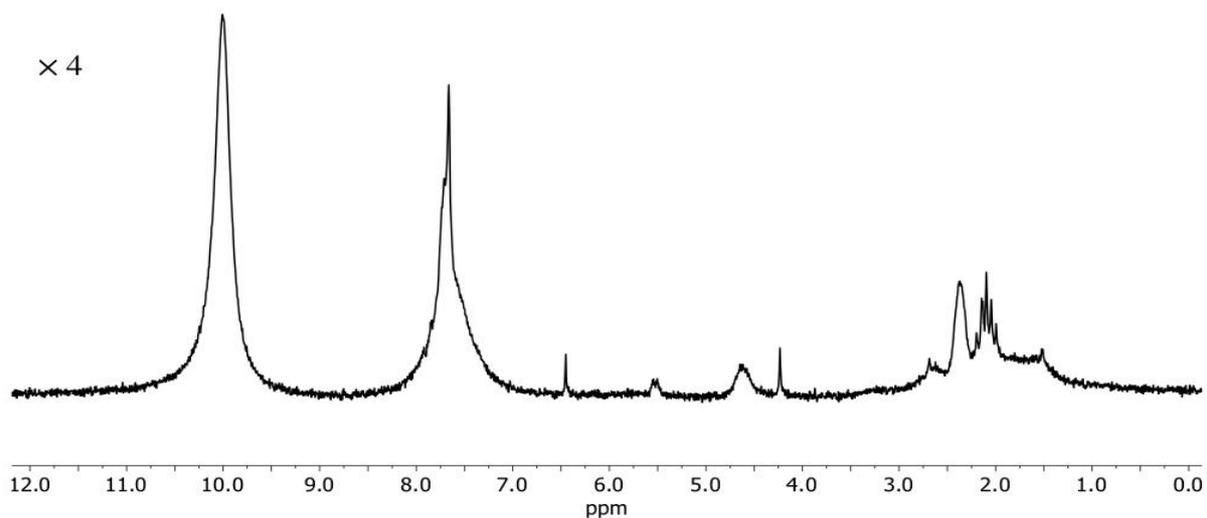

S19: Thermal $^1$H NMR spectrum recorded 10 minutes after the acquisition of S18. Compared to the corresponding hyperpolarized spectrum, this spectrum is enlarged by a factor of 4.

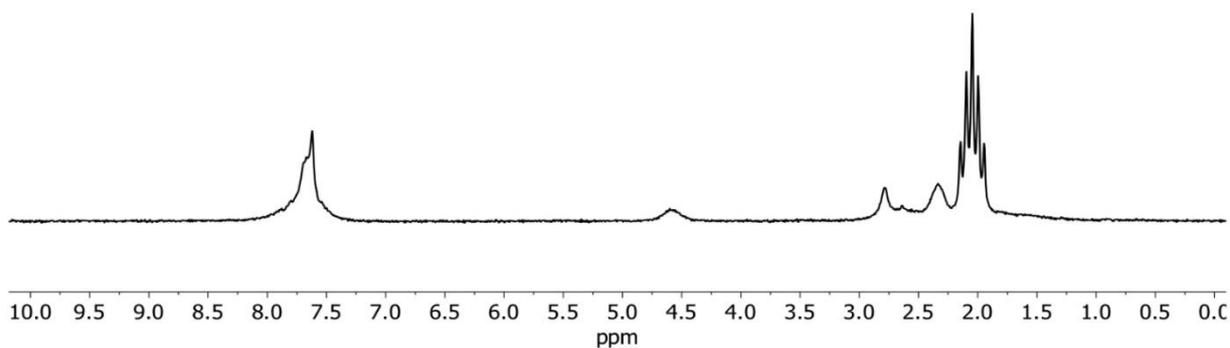

S20: Thermal $^1$H NMR spectrum of propargyl glycine in acetone-$d_6$ with 5.5 mM [Rh(dppb)(COD)]BF4.

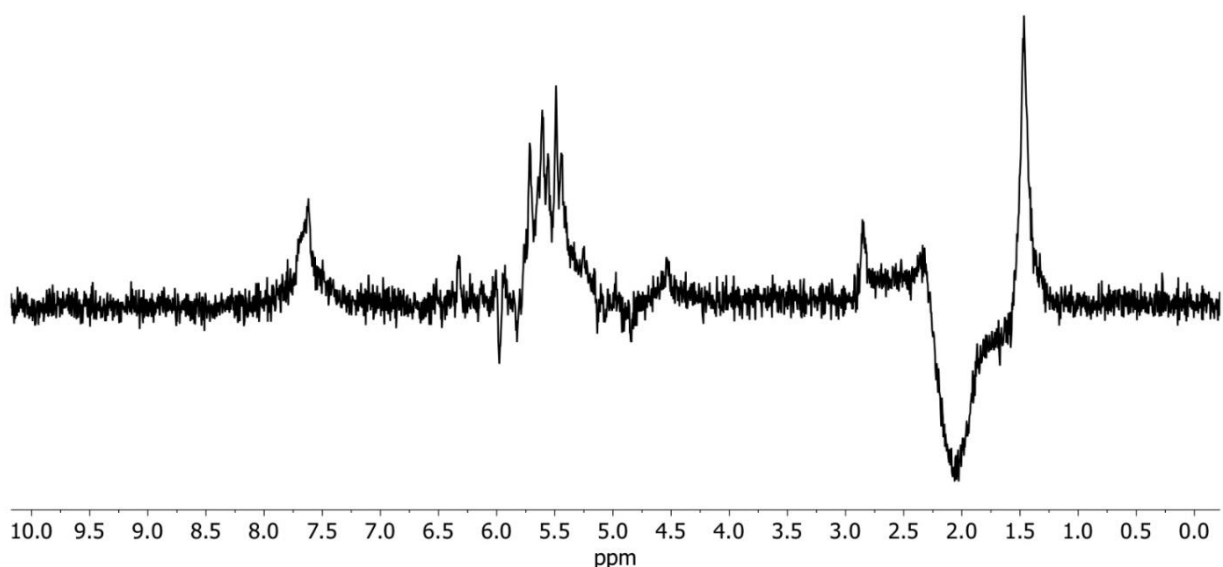

S21: Hyperpolarized $^1$H NMR spectrum of parahydrogenated propargyl glycine in acetone-$d_6$.

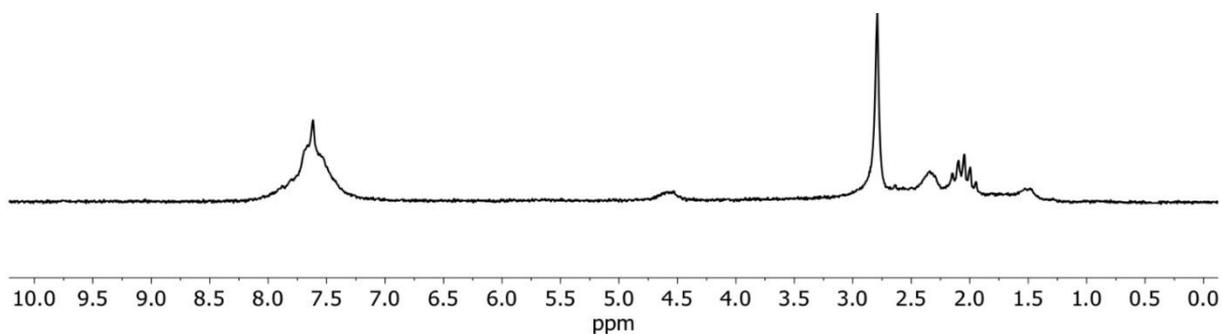
S22: Thermal ¹H NMR spectrum recorded 10 minutes after the acquisition of S21.

**Parahydrogenation in dichloromethane-d$_2$:**

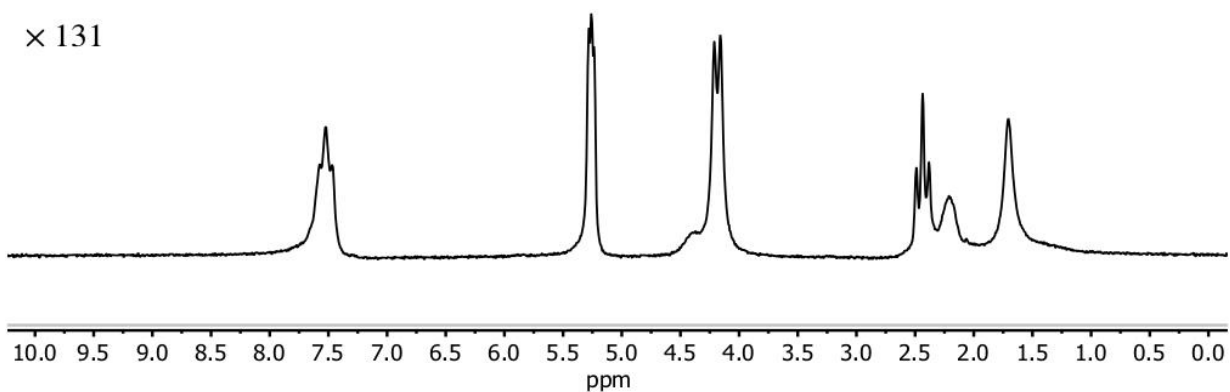
S23: Thermal ¹H NMR spectrum of 85 mM propargyl alcohol in dichloromethane-d$_2$ with 5.5 mM [Rh(dppb)(COD)]BF$_4$. Compared to the corresponding hyperpolarized spectrum, this spectrum is enlarged by a factor of 131.

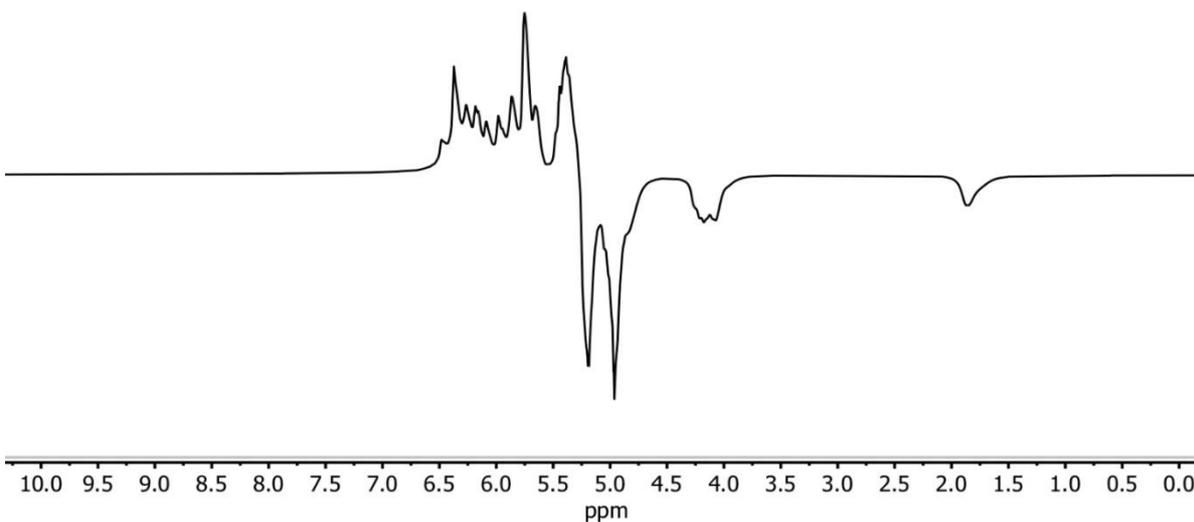
S24: Hyperpolarized ¹H NMR spectrum of parahydrogenated propargyl alcohol in dichloromethan-d2.

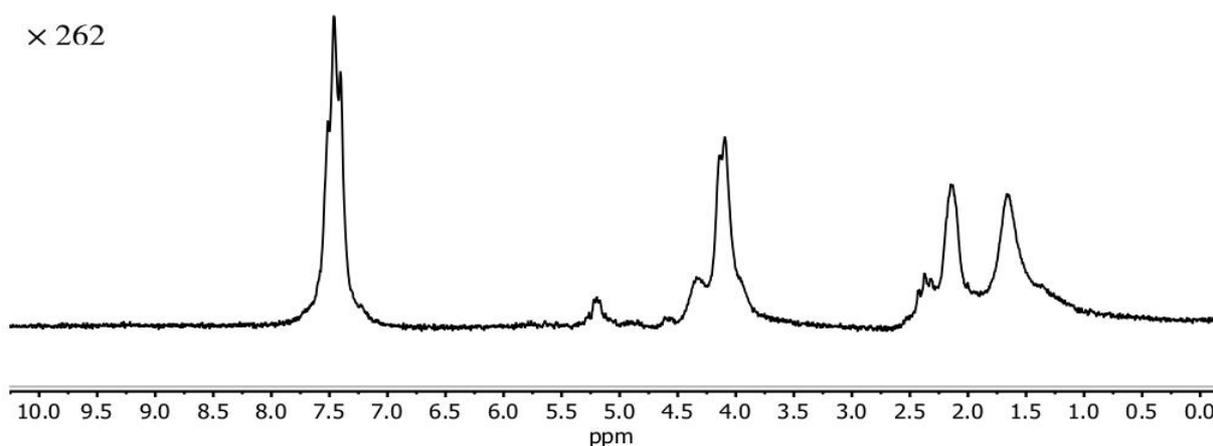

S25: Thermal ¹H NMR spectrum recorded 10 minutes after the acquisition of S24. Compared to the corresponding hyperpolarized spectrum, this spectrum is enlarged by a factor of 262.

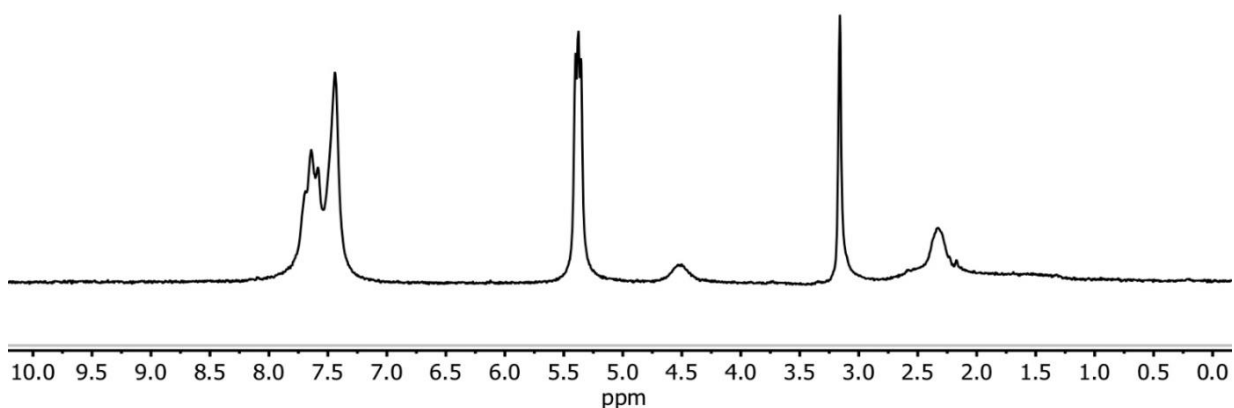

S26: Thermal ¹H NMR spectrum of 85 mM propiolic acid in dichloromethane-d$_2$ with 5.5 mM [Rh(dppb)(COD)]BF$_4$. Compared to the corresponding hyperpolarized spectrum, this spectrum is not enlarged.

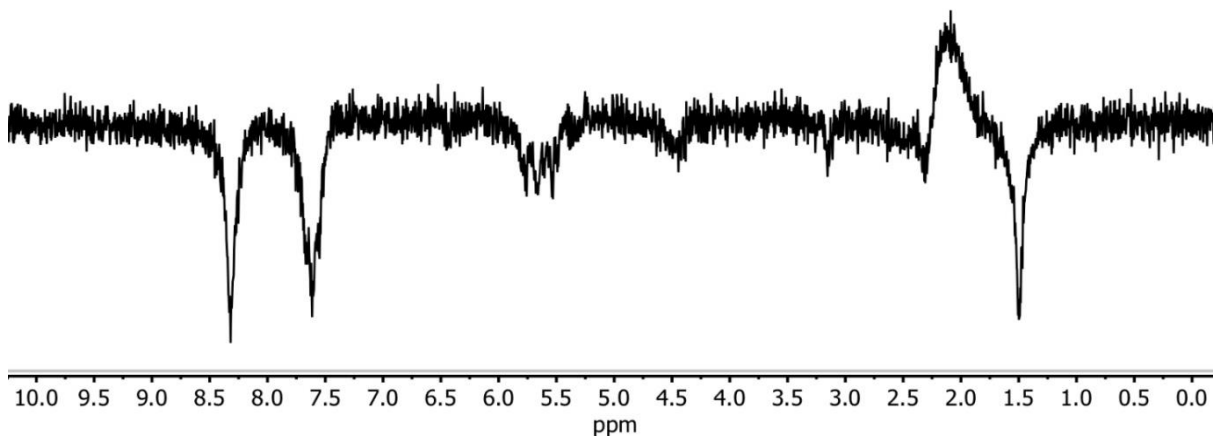

S27: Hyperpolarized ¹H NMR spectrum of parahydrogenated propiolic acid in dichloromethan-d$_2$.

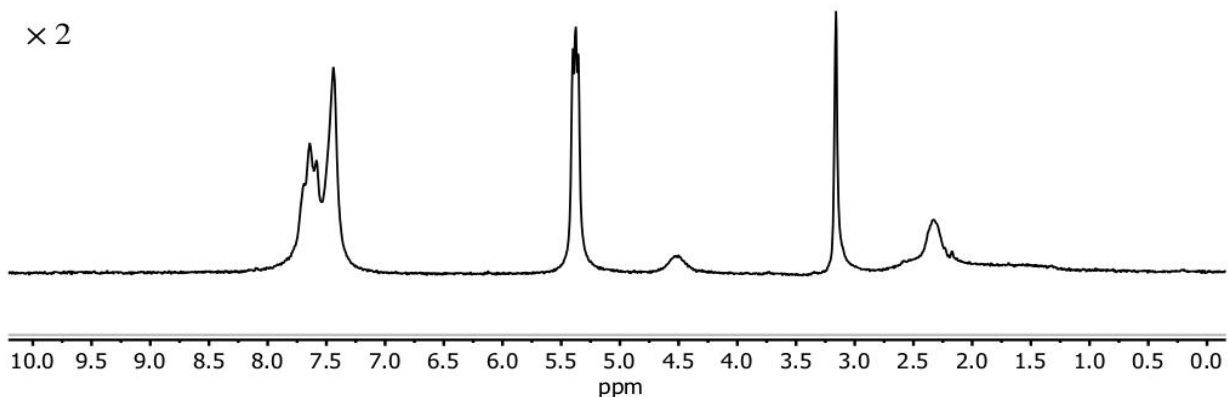

S28: Thermal ¹H NMR spectrum recorded 10 minutes after the acquisition of S27. Compared to the corresponding hyperpolarized spectrum, this spectrum is enlarged by a factor of 2.

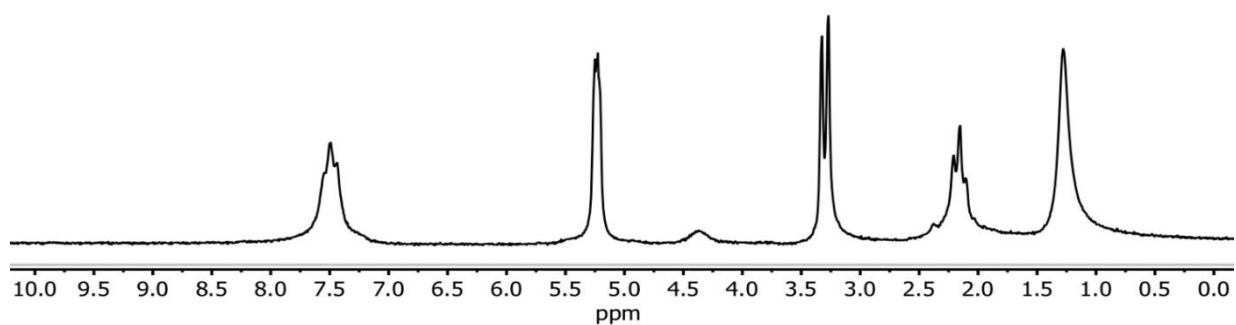

S29: Thermal ¹H NMR spectrum of 85 mM propargyl amine in dichloromethane-d2 with 5.5 mM [Rh(dppb)(COD)]BF₄. Compared to the corresponding hyperpolarized spectrum, this spectrum is enlarged by a factor of 131.

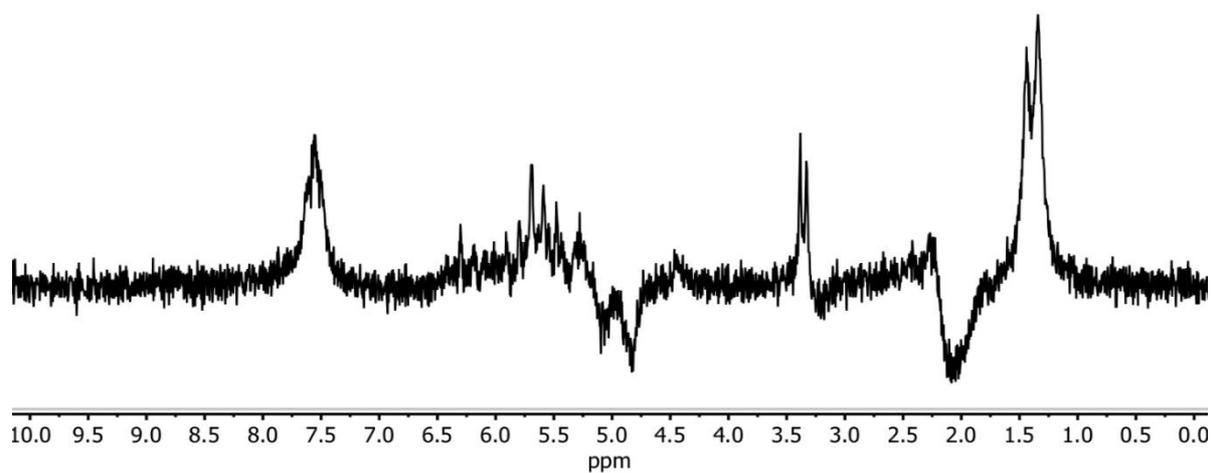

S30: Hyperpolarized ¹H NMR spectrum of parahydrogenated propargyl amine in dichloromethan-d2.

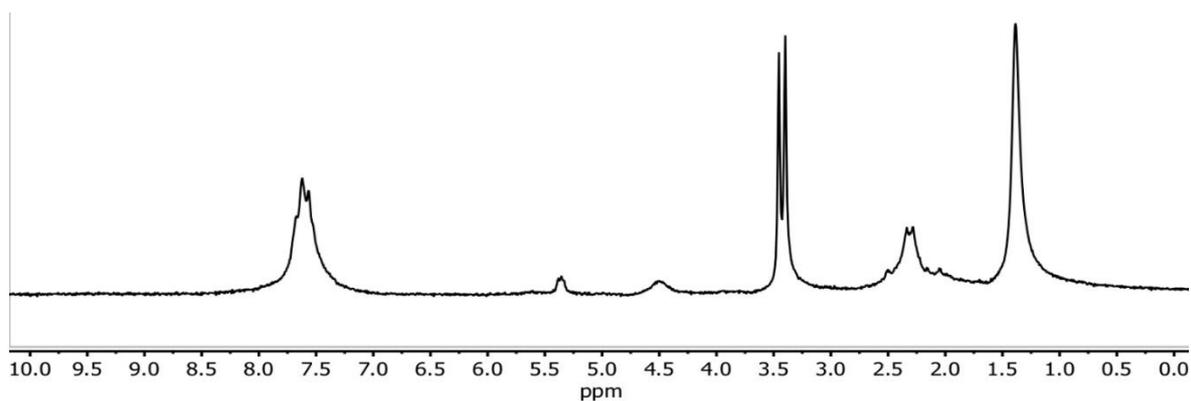

S31: Thermal ¹H NMR spectrum recorded 10 minutes after the acquisition of S30.

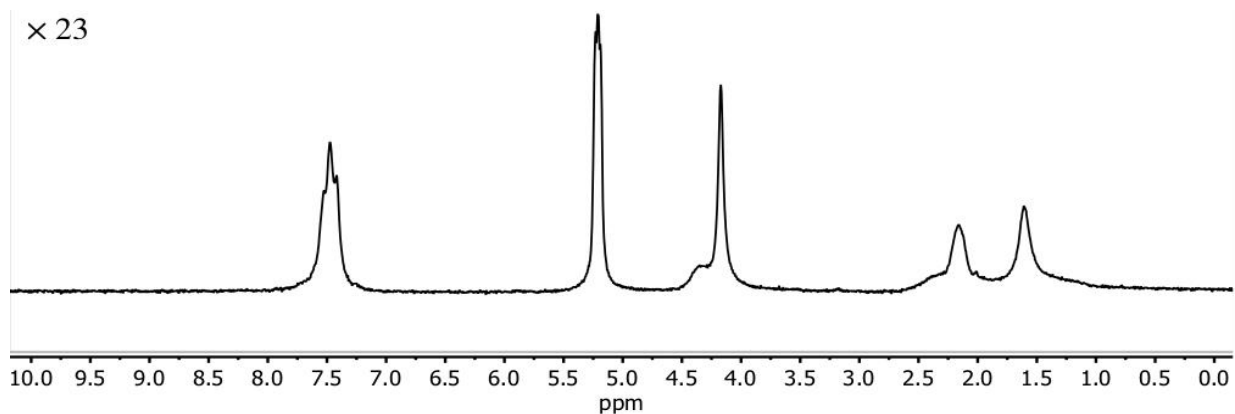

S32: Thermal ¹H NMR spectrum of 85 mM butyne-1,4-diol in dichloromethane-d$_2$ with 5.5 mM [Rh(dppb)(COD)]BF$_4$. Compared to the corresponding hyperpolarized spectrum, this spectrum is enlarged by a factor of 23.

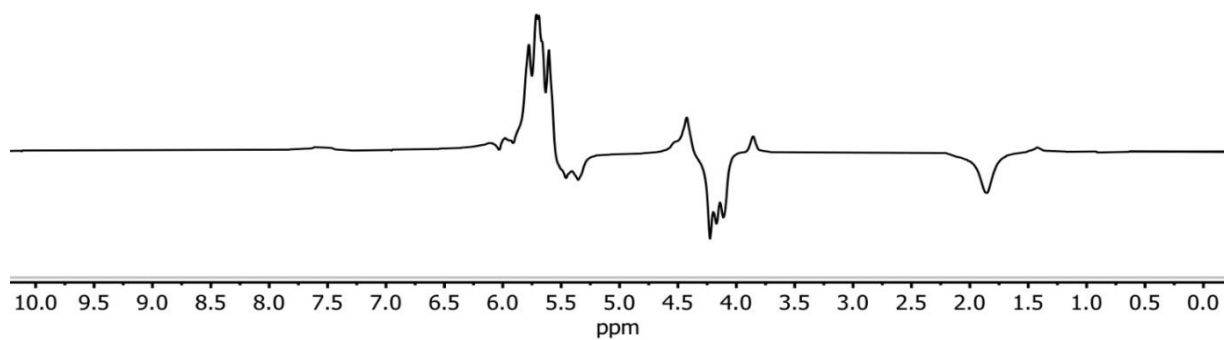

S33: Hyperpolarized ¹H NMR spectrum of parahydrogenated butyne-1,4-diol in dichloromethan-d$_2$.

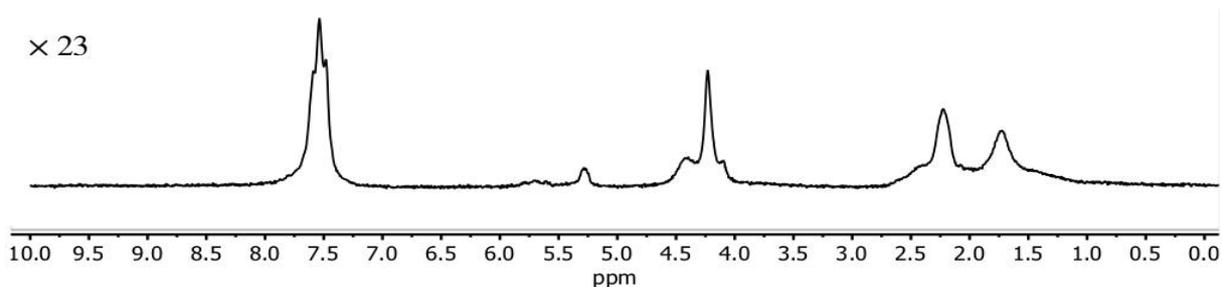

S34: Thermal ¹H NMR spectrum recorded 10 minutes after the acquisition of S33. Compared to the corresponding hyperpolarized spectrum, this spectrum is enlarged by a factor of 23.

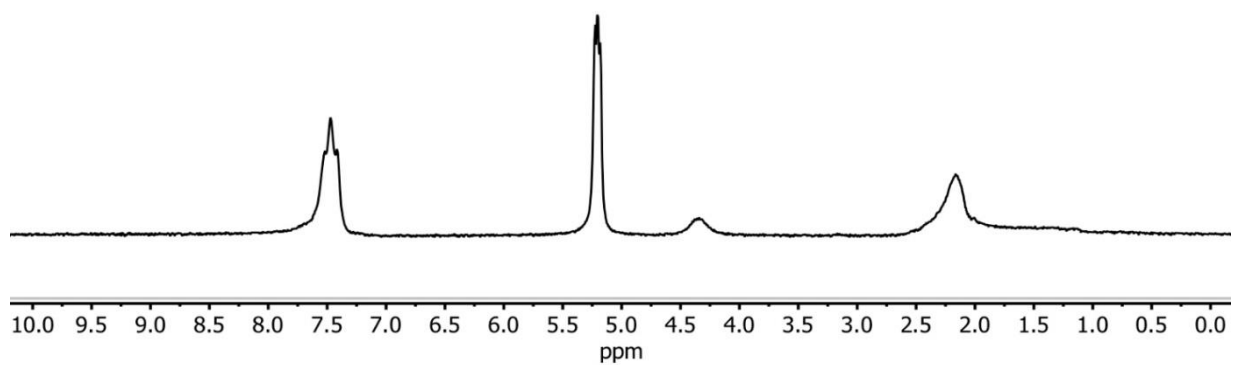

S35: Thermal ¹H NMR spectrum of 85 mM acetylendicarboxylic acid in dichloromethane-$d_2$ with 5.5 mM [Rh(dppb)(COD)]BF4. Compared to the corresponding hyperpolarized spectrum, this spectrum is not enlarged.

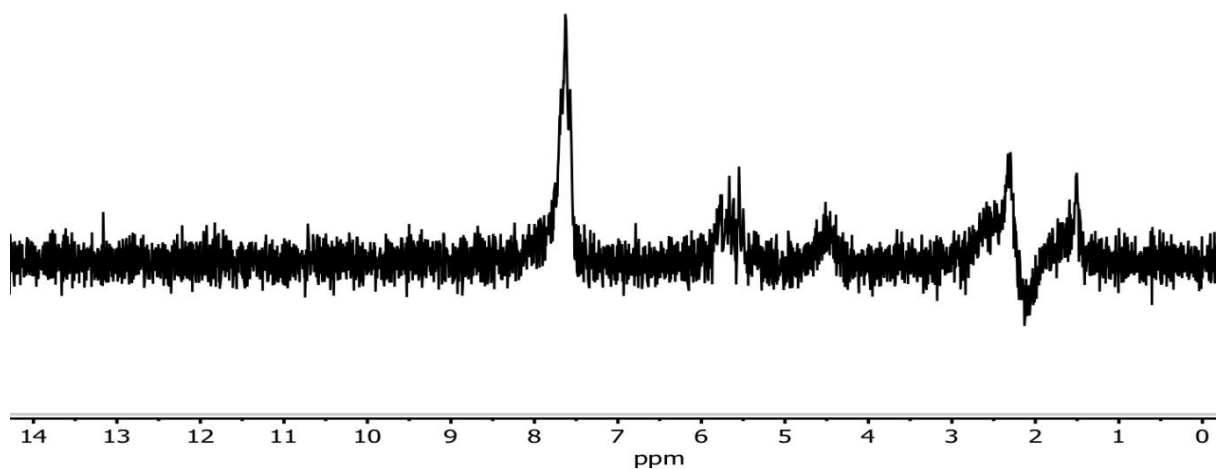

S36: Hyperpolarized ¹H NMR spectrum of parahydrogenated acetylene dicarboxylic acid in dichloromethan-$d_2$.

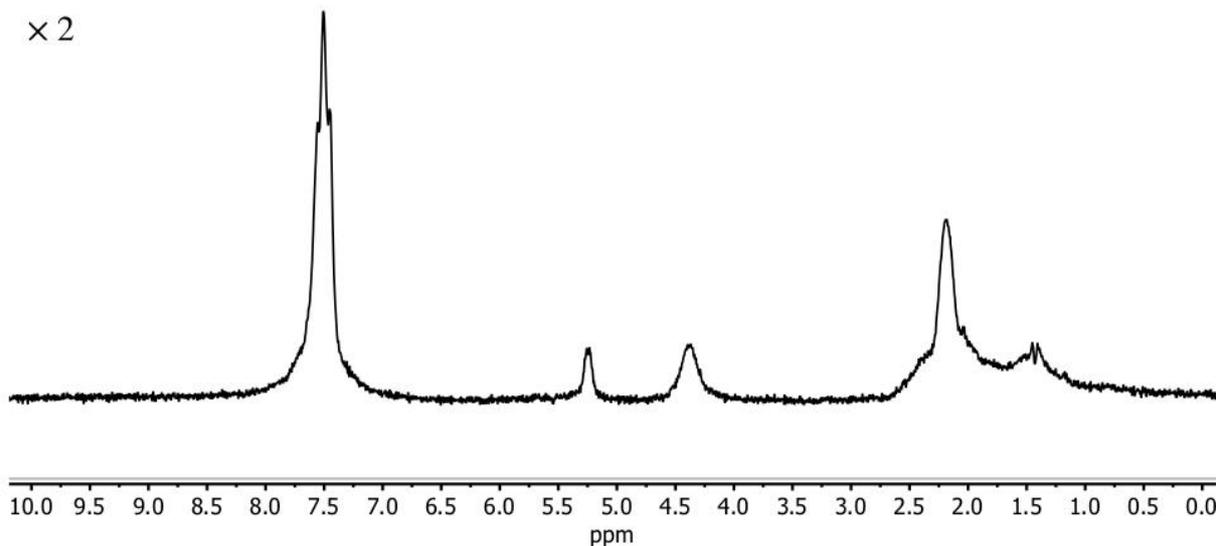

S37: Thermal ¹H NMR spectrum recorded 10 minutes after the acquisition of S36. Compared to the corresponding hyperpolarized spectrum, this spectrum is enlarged by a factor of 2.

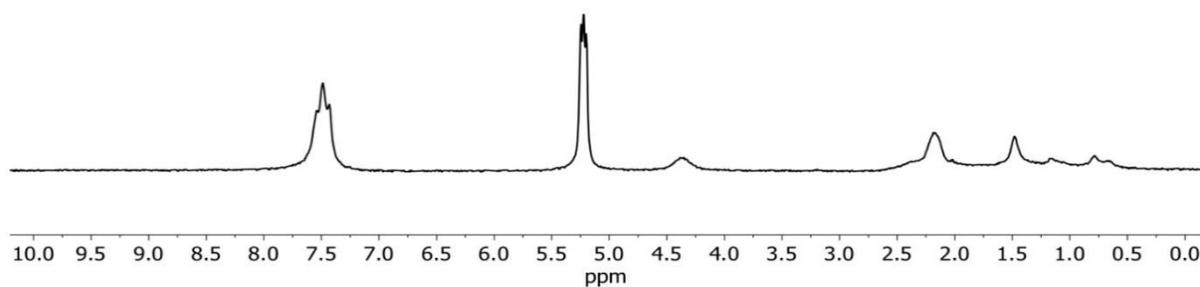

S38: Thermal ¹H NMR spectrum of propargyl glycine in dichloromethane-d$_2$ with 5.5 mM [Rh(dppb)(COD)]BF$_4$. Compared to the corresponding hyperpolarized spectrum, this spectrum is not enlarged.

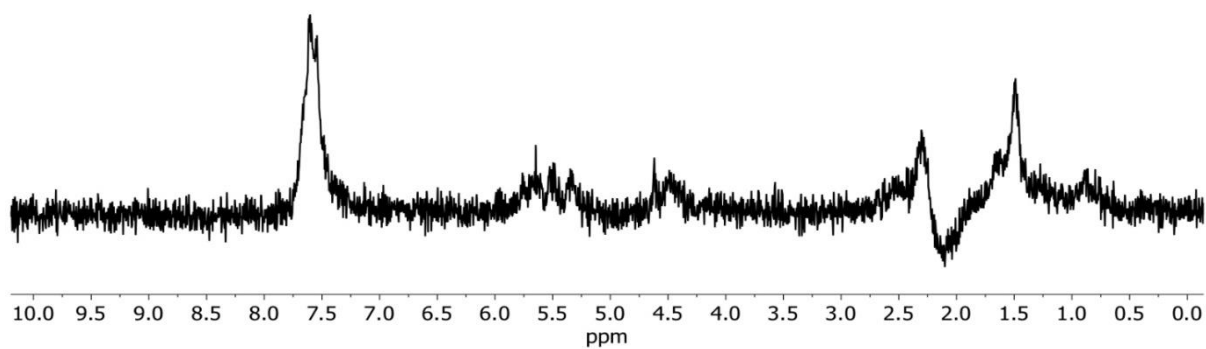

S39: Hyperpolarized ¹H NMR spectrum of parahydrogenated propargyl glycine in dichloromethan-d2.

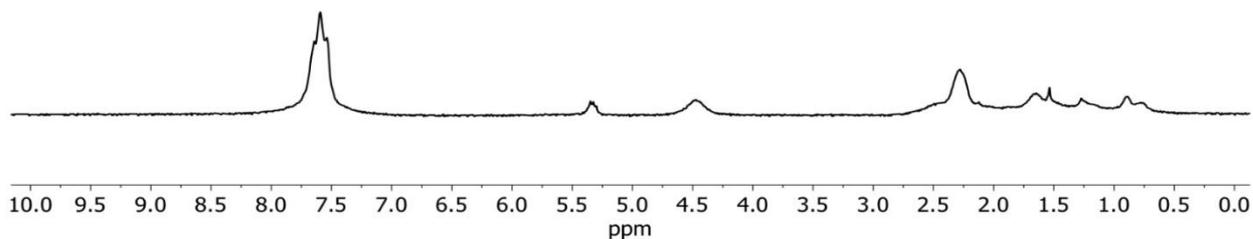

S40: Thermal ¹H NMR spectrum recorded 10 minutes after the acquisition of S39.

**Parahydrogenation in DMSO-d6:**

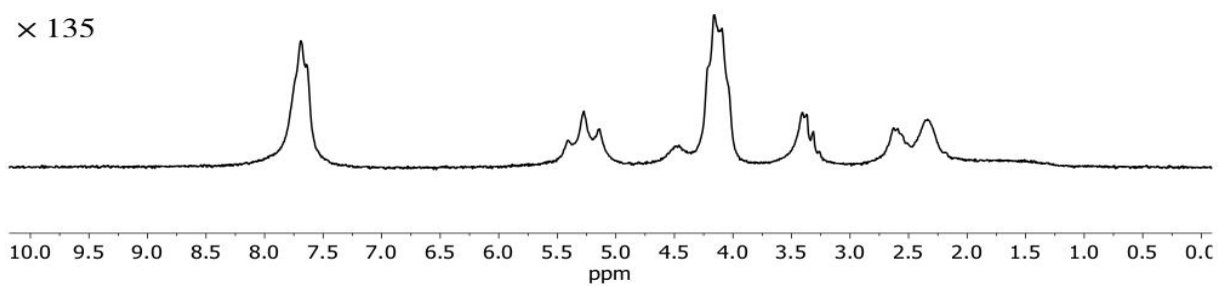

S41: Thermal ¹H NMR spectrum of 85 mM propargyl alcohol in dimethyle sulfoxide-d2 with 5.5 mM [Rh(dppb)(COD)]BF4. Compared to the corresponding hyperpolarized spectrum, this spectrum is enlarged by a factor of 135.

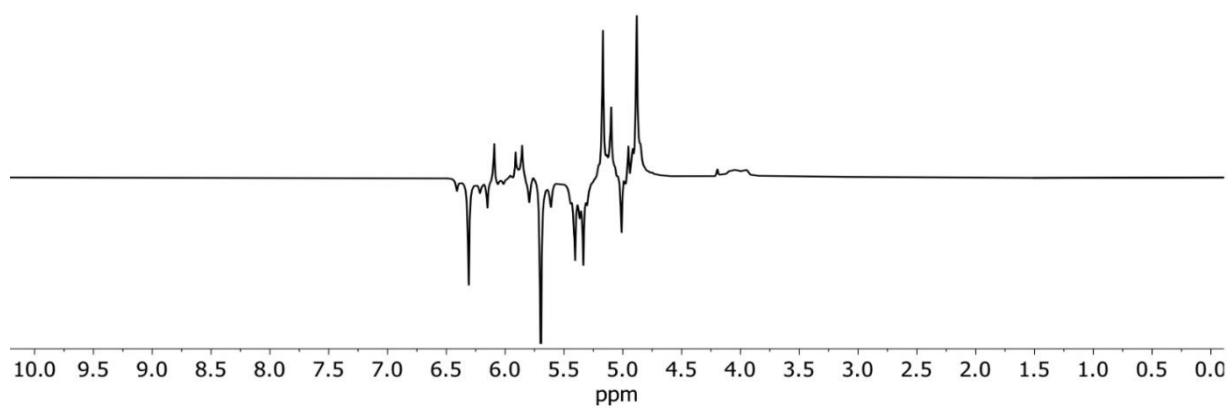

S42: Hyperpolarized ¹H NMR spectrum of parahydrogenated propargyl alcohol in dimethyl sulfoxide-d6.

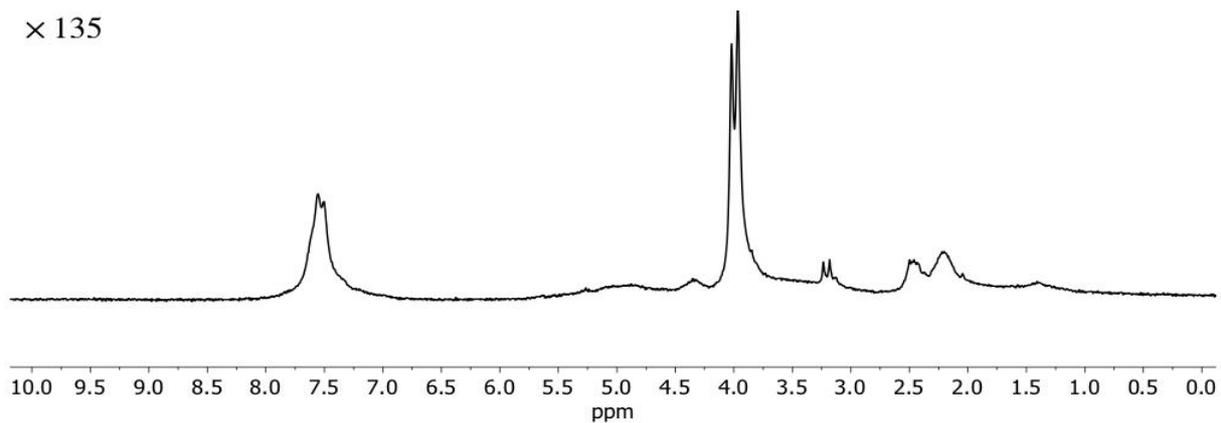

S43: Thermal ¹H NMR spectrum recorded 10 minutes after the acquisition of S44. Compared to the corresponding hyperpolarized spectrum, this spectrum is enlarged by a factor of 135.

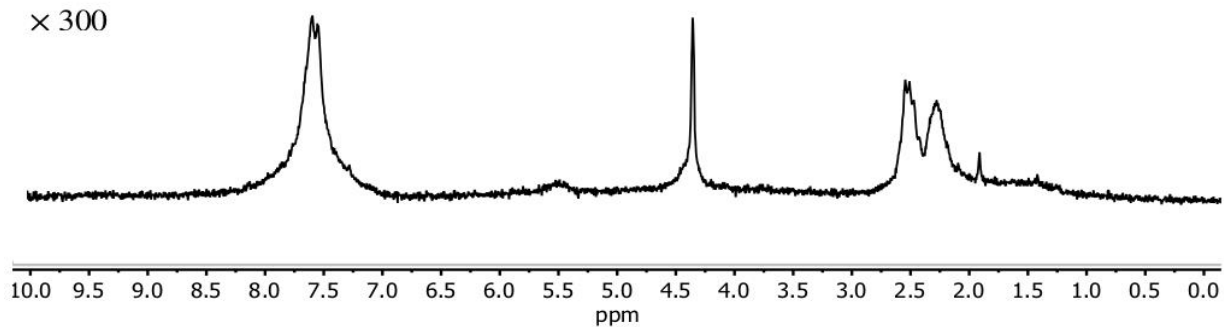

S44: Thermal ¹H NMR spectrum of 85 mM propiolic acid in dimethyl sulfoxide-d2 with 5.5 mM [Rh(dppb)(COD)]BF4. Compared to the corresponding hyperpolarized spectrum, this spectrum is enlarged by a factor of 300.

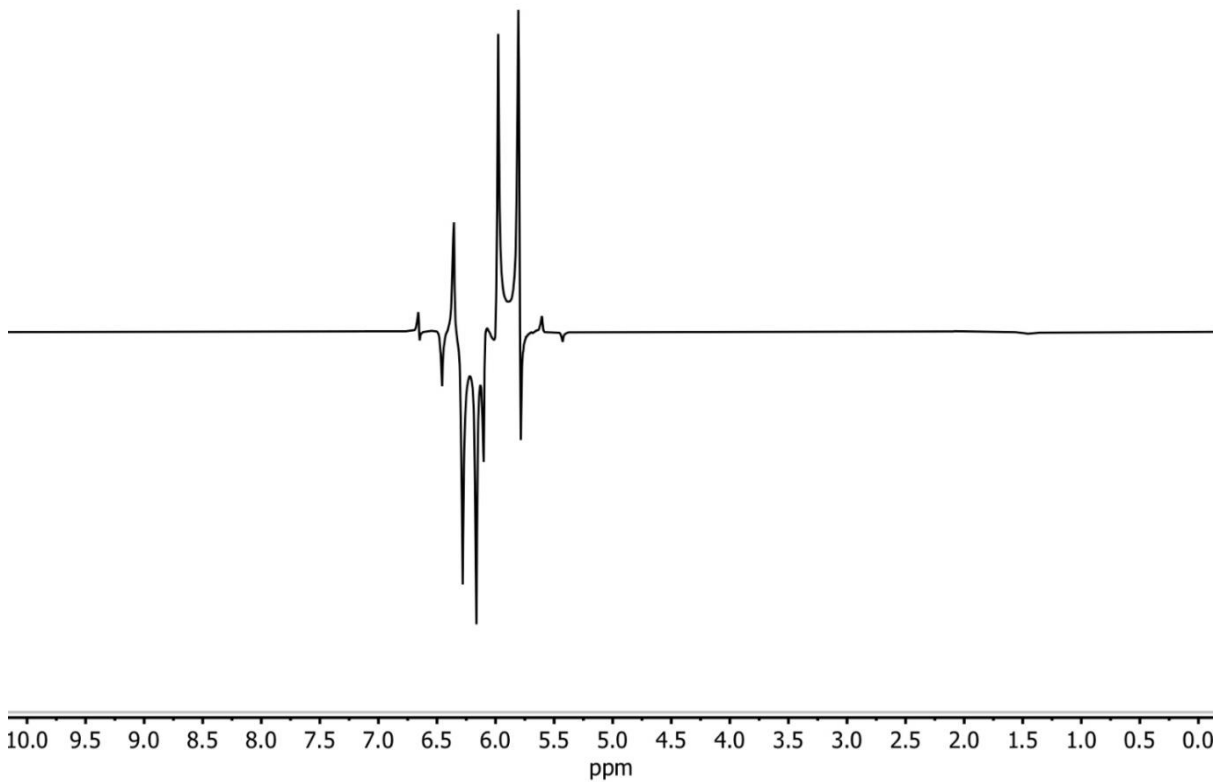

S45: Hyperpolarized ¹H NMR spectrum of parahydrogenated propiolic acid in dimethyl sulfoxide-d6.

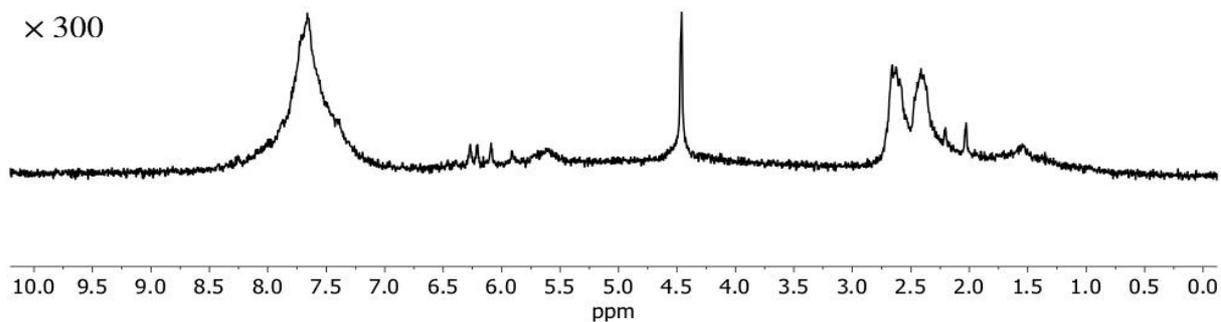

S46: Thermal ¹H NMR spectrum recorded 10 minutes after the acquisition of S45. Compared to the corresponding hyperpolarized spectrum, this spectrum is enlarged by a factor of 300.

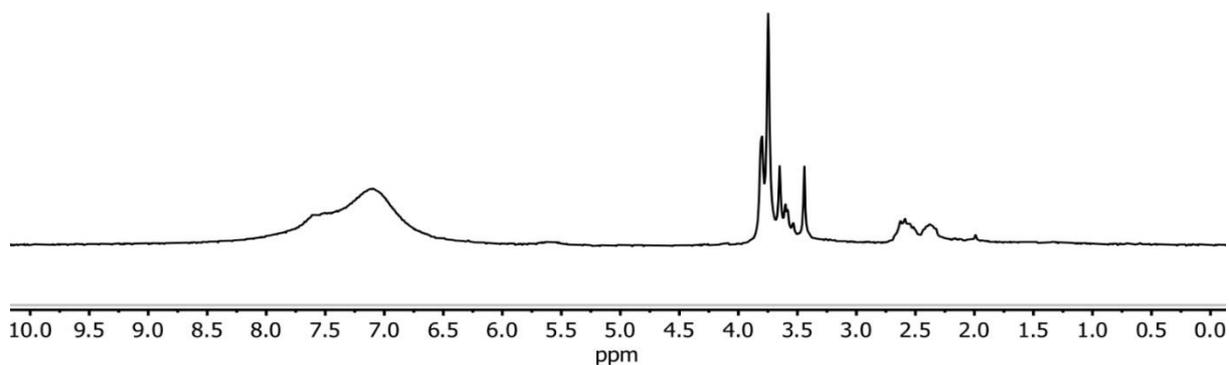

S47: Thermal ¹H NMR spectrum of 85 mM propargyl amine in dimethyl sulfoxide-d2 with 5.5 mM [Rh(dppb)(COD)]BF4. Compared to the corresponding hyperpolarized spectrum, this spectrum is not enlarged.

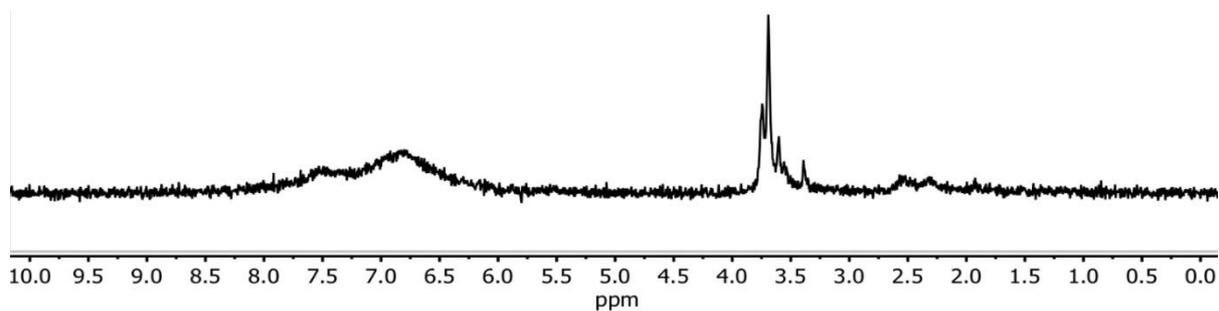

S48: ¹H NMR spectrum of parahydrogenated propargyl amine in dimethyl sulfoxide-d6.

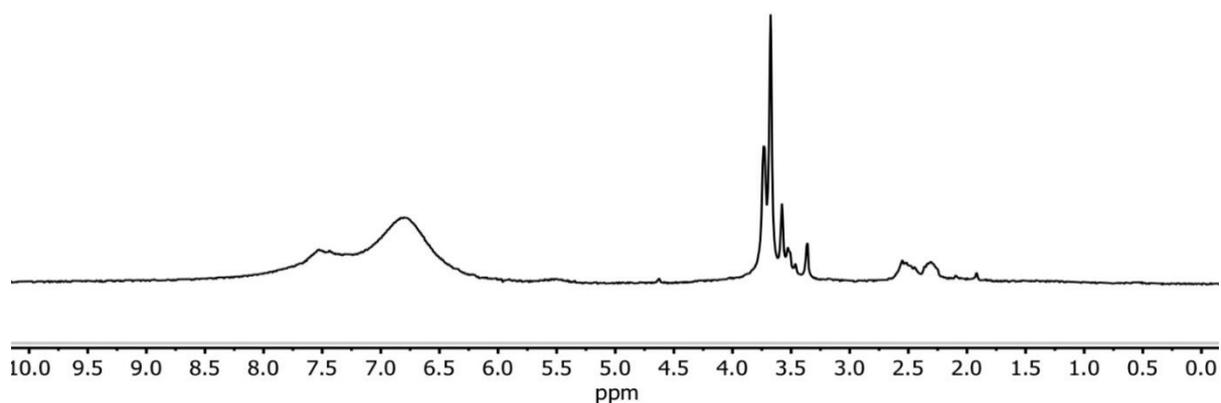

S49: Thermal ¹H NMR spectrum recorded 10 minutes after the acquisition of S48.

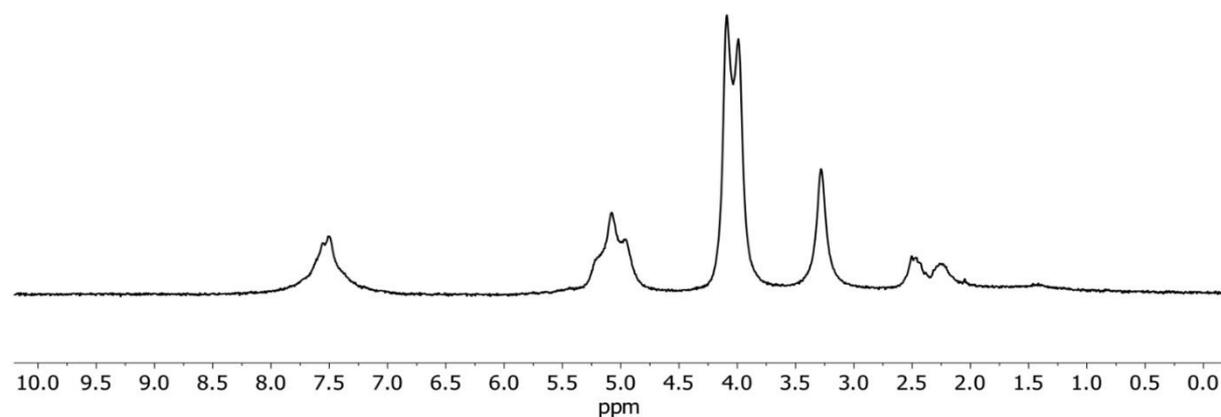

S50: Thermal ¹H NMR spectrum of 85 mM butyne-1,4-diol in dimethyl sulfoxide-d2 with 5.5 mM [Rh(dppb)(COD)]BF4. Compared to the corresponding hyperpolarized spectrum, this spectrum is enlarged by a factor of 135.

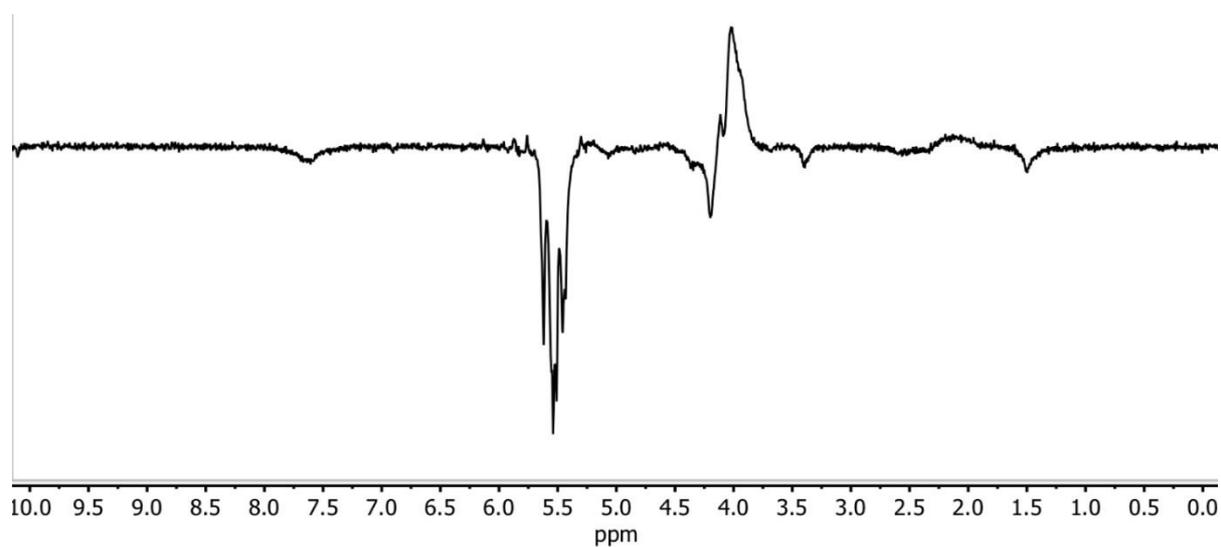

S51: Hyperpolarized ¹H NMR spectrum of parahydrogenated butyne-1,4-diol in dimethyl sulfoxide-d6.

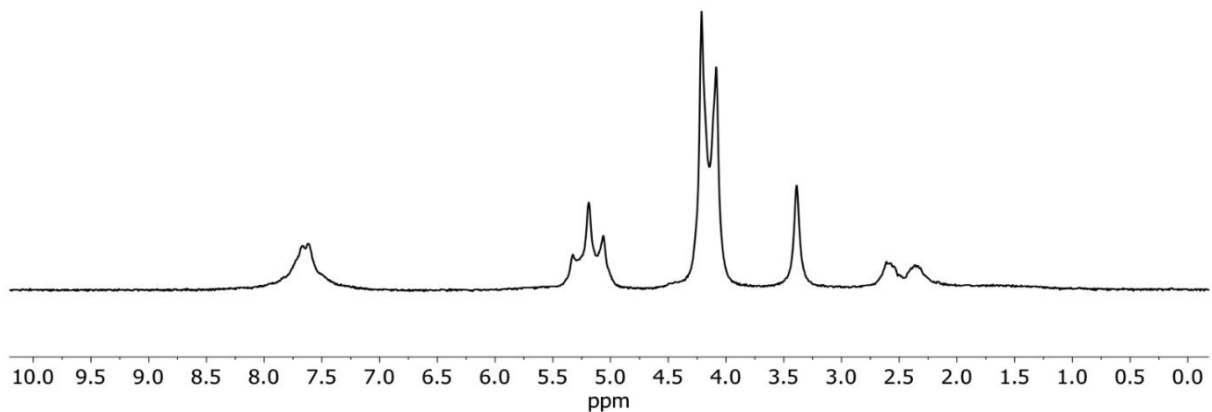

S52: Thermal ¹H NMR spectrum recorded 10 minutes after the acquisition of S51.

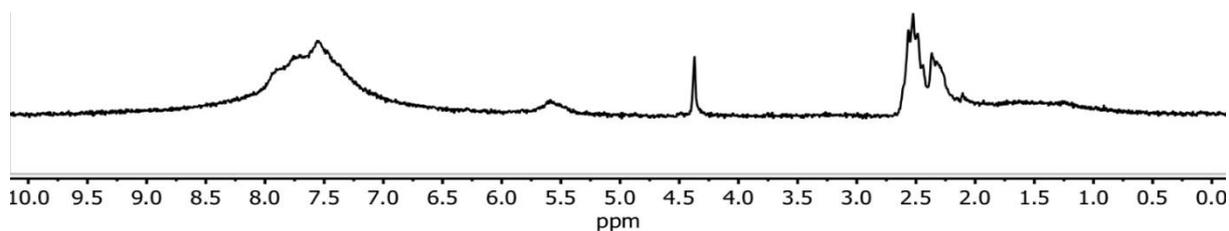

S53: Thermal ¹H NMR spectrum of acetylenedicarboxylic acid in dimethyl sulfoxide-d2 with 5.5 mM [Rh(dppb)(COD)]BF4. Compared to the corresponding hyperpolarized spectrum, this spectrum is not enlarged.

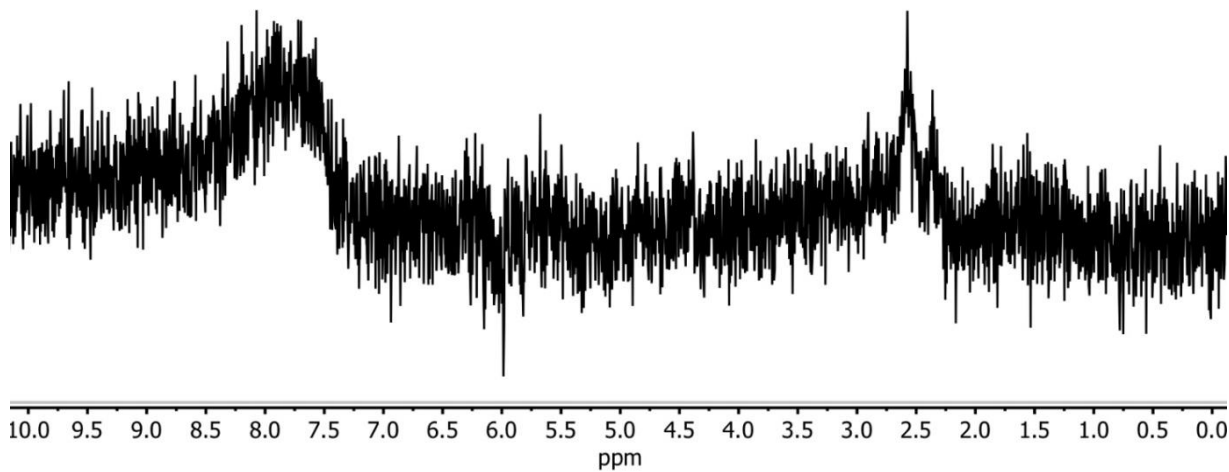

S54: ¹H NMR spectrum of parahydrogenated acetylenedicarboxylic acid in dimethyl sulfoxide-d6.

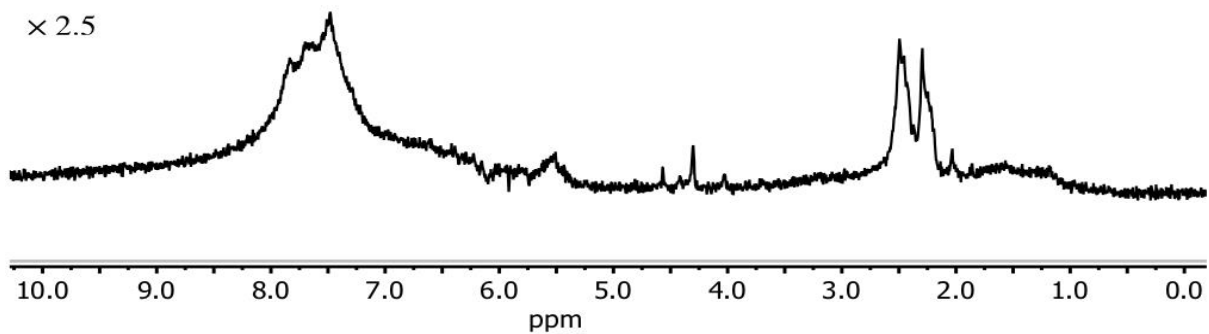

S55: Thermal ¹H NMR spectrum recorded 10 minutes after the acquisition of S54. Compared to the corresponding hyperpolarized spectrum, this spectrum is enlarged by a factor of 2.5.

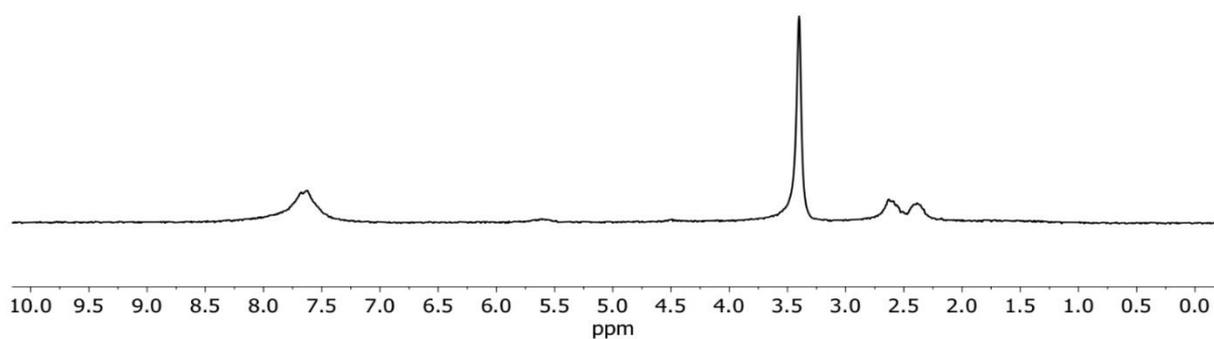

S56: Thermal ¹H NMR spectrum of propargyl glycine in dimethyl sulfoxide-$d_2$ with 5.5 mM [Rh(dppb)(COD)]BF$_4$. Compared to the corresponding hyperpolarized spectrum, this spectrum is enlarged by a factor of 135.

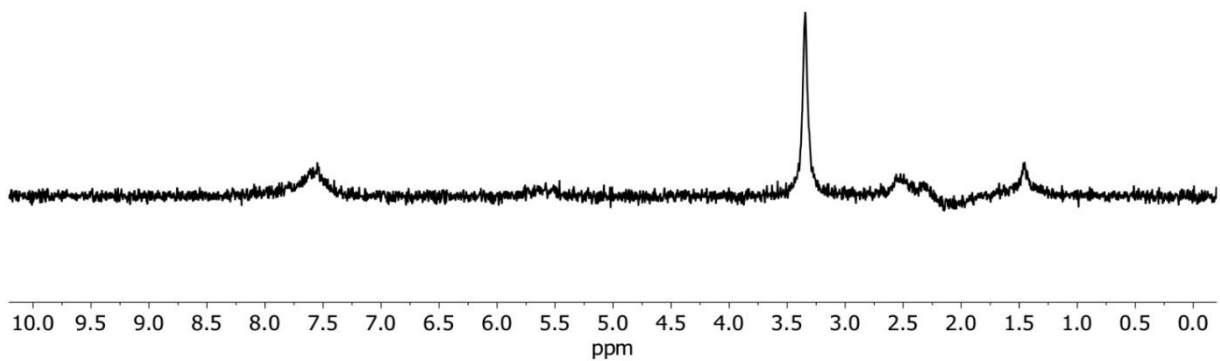

S57: ¹H NMR spectrum of parahydrogenated propargyl glycine in dimethyl sulfoxide-$d_6$.

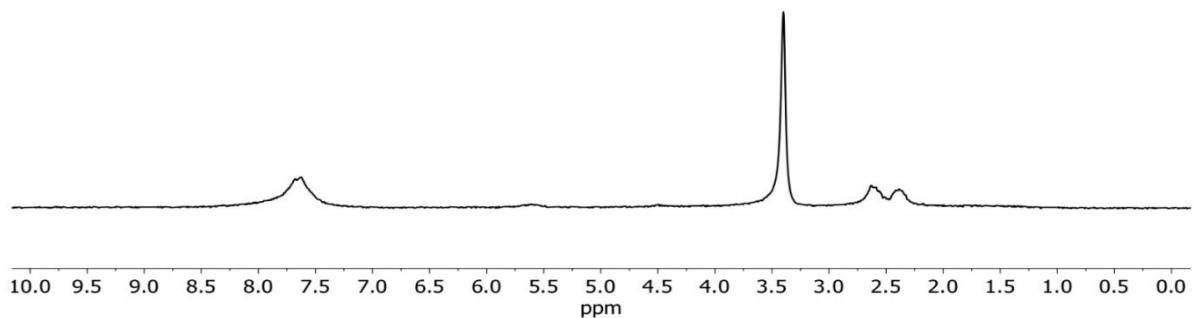

S58: Thermal ¹H NMR spectrum recorded 10 minutes after the acquisition of S57.

## 5. Lactate / Lactic acid hyperpolarized by PHIP-X

Lactic acid was hyperpolarized using PHIP-X. To this end, pH$_2$ was bubbled for 60 seconds at atmospheric pressure and 6 mT external magnetic field through a solution containing 29 mM of lactic acid, 50 mM propargyl alcohol, and 7 mM of catalyst Rh(dppb)(COD)]BF$_4$ in acetone-d6. The polarization of both the labile OH proton and the labile COOH proton were difficult to estimate because they were distributed over a large range in the ppm scale and overlapped with many signals. Therefore, we show that proton no. 13 of figure S59 was hyperpolarized. In order to identify the corresponding signal in PHIP-X spectrum we start with a thermal spectrum of 29 mM of lactic acid in acetone-d6 (S60).

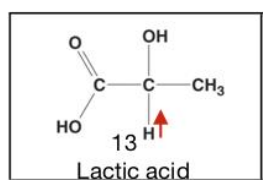

S59: Structure of lactic acid. It was found that proton no. 13 was hyperpolarized, which is denoted by the red arrow.

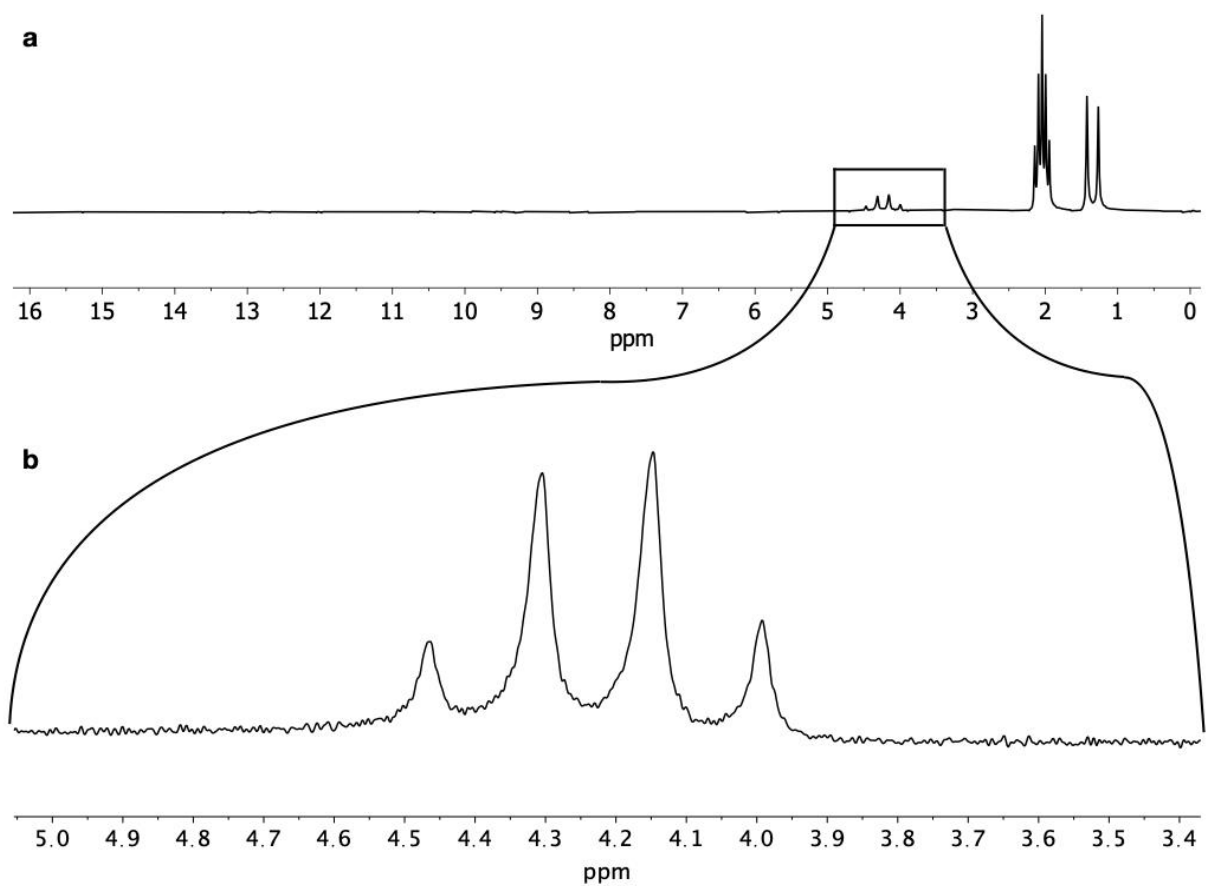

S60: (a) Thermal $^1$H NMR spectrum of 29 mM lactic acid in acetone-d6 from 0.0 ppm - 16 ppm. NMR signals of the COOH and the OH peak are not visible in this scale because they are broadly smeared and have therefore a low amplitude. The four peaks between 4.0 - 4.5 ppm are generated by proton no. 13 shown in figure S59. A zoom to this region (b) shows clearly at which ppm the nucleus absorbs energy from the pulse sequence. Of special importance are the peaks at 4.3 and 4.44 ppm because they are visible in PHIP-X NMR spectra.

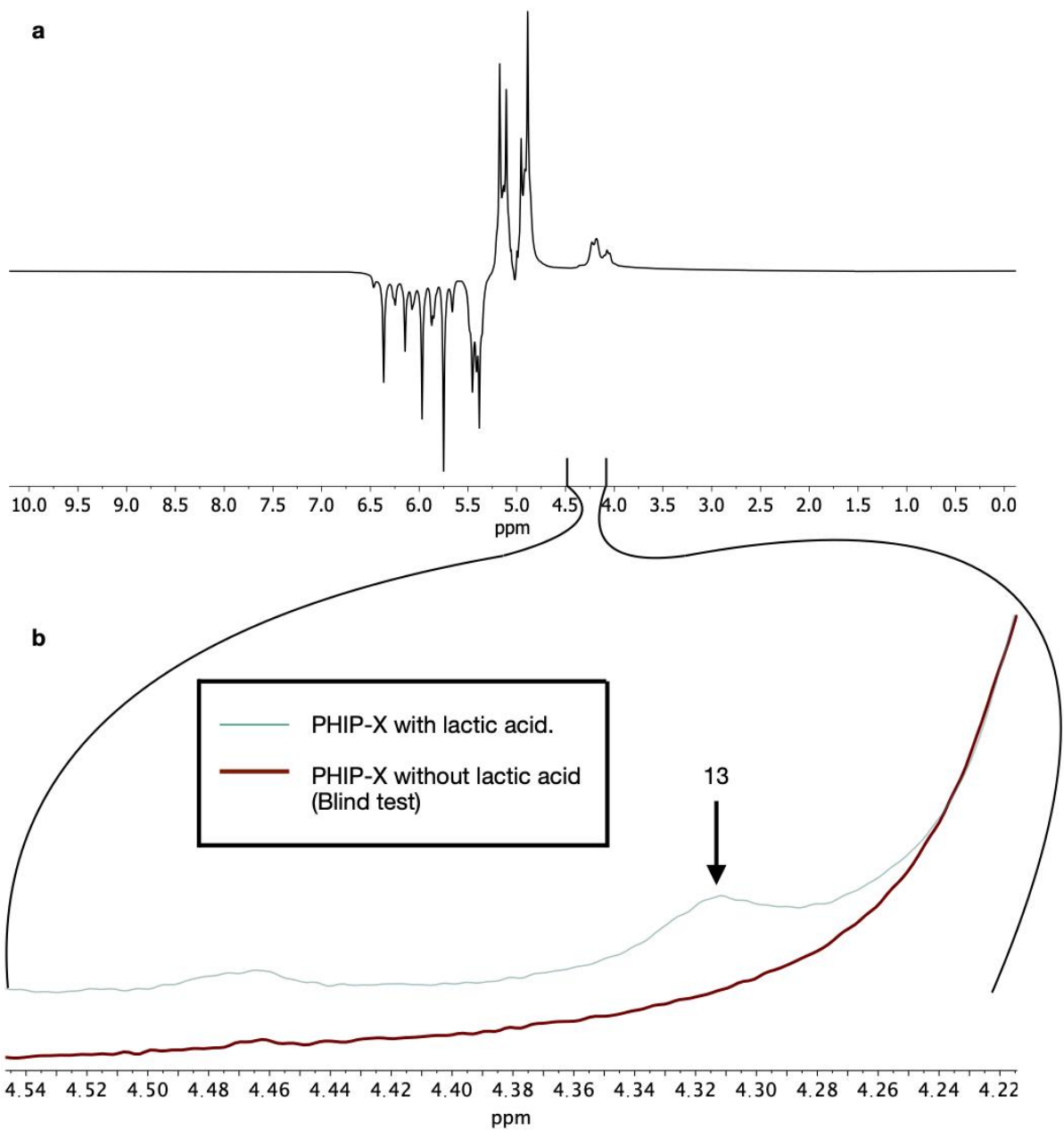

S61: (a) PHIP-X ¹H NMR spectrum with 29 mM lactic acid from 0.0 ppm to 10.0 ppm. A zoom to the region from 4.22 ppm - 4.54 ppm with a comparison of a blind test (b) clearly shows the hyperpolarized nuclear spin no. 13 from lactic acid. The chemical shifts at 4.31 and 4.47 for this proton are in agreement with the thermal spectrum shown in fig. S60. The blind test consisted of the same solution but without lactic acid. The procedure for parahydrogenation and signal acquisition was the same for both experiments, PHIP-X with lactic acid and the blind test.

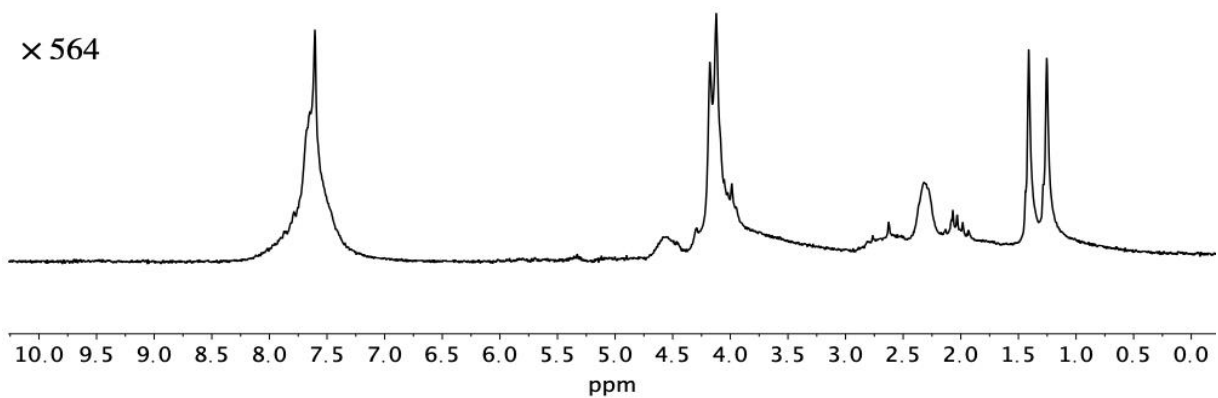

S62: Thermal ¹H NMR spectrum recorded 10 minutes after the PHIP-X experiment whose NMR spectrum is shown in fig. S61. Compared to the corresponding hyperpolarized spectrum, fig. S61a, this spectrum is enlarged by a factor of 564.

6. Pyruvic acid hyperpolarized by PHIP-X

A relatively high concentration of 120 mM of pyruvic acid was required to identify the NMR signal of the hyperpolarized COOH proton of pyruvic acid (fig. S63). This is because for lower concentrations relatively larger amounts of the COOH protons become dissociated due to the interaction with water. Furthermore, the NMR signal of the COOH proton becomes extremely broad for lower concentrationsas a result of the interaction with other labile protons from water, propargyl alcohol, and allyl alcohol. Due to the high concentration of 120 mM, the polarization obtained is lower because the polarization is either shared by many more molecules or only a small fraction effectively gets polarized. This fraction, however, cannot be distinguished in the NMR spectra. It was observed that the presence of pyruvic acid disturbs the hydrogenation reaction of 1a to 1b such that fewer polarized transfer molecules are produced. In order to investigate the hypothesis that this problem can be solved, we evaluated experiments in which the pyruvic acid was injected soon after the parahydrogenation of 1a to 1b. Indeed, much stronger (about 10-fold) NMR signals of 1b were observed, confirming that this problem can be handled that way. Note that this 10-fold stronger signal is still about 3 times smaller than the blind test shown in fig. 4 of the main manuscript. A major reason for this is given by relaxation effects due to the longer time which was required for the procedure when pyruvic acid was injected after the hydrogenation. However, in this case the NMR signal of the COOH proton also becomes strongly broadened and overlaps with signals of 1b. Hence, a quantification of the polarization for this proton is difficult here and may contain an error that is difficult to estimate. Nevertheless, higher polarizations are estimated.

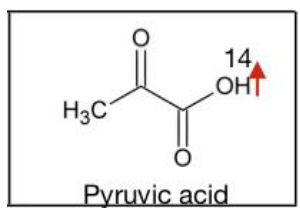

S63: Structure of pyruvic acid with number for the proton which was investigated in PHIP-X experiments. The red arrow indicates that this proton was hyperpolarized.

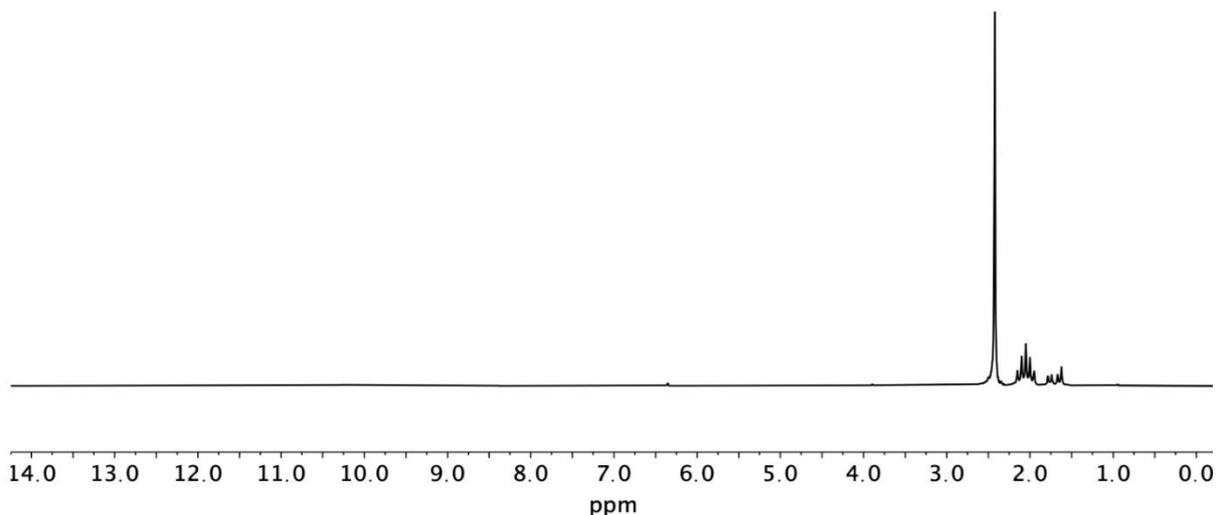

S64: Thermal ¹H NMR spectrum of 120 mM pyruvic acid in acetone-d6. Most of the NMR signal of the COOH proton is located between 8.5 ppm and 12.0 ppm. Due to the very large width of the signal, the height is lowered accordingly. Therefore, we show a 27-fold amplification of this spectrum in fig. S65.

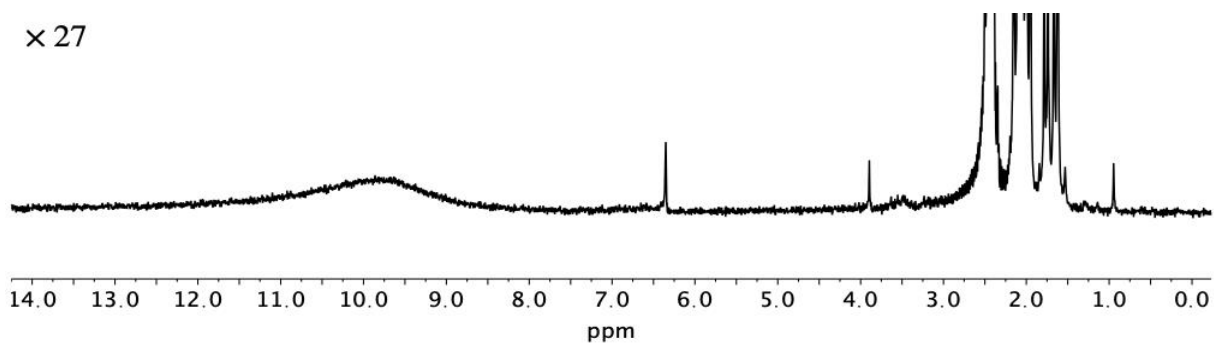

S65: 27-fold zoom of S64. The COOH peak is centered at about 9.8 ppm. The interaction with other labile protons from water, propargyl alcohol, or allyl alcohol broadens the peaks and shifts the peak towards lower ppm values.

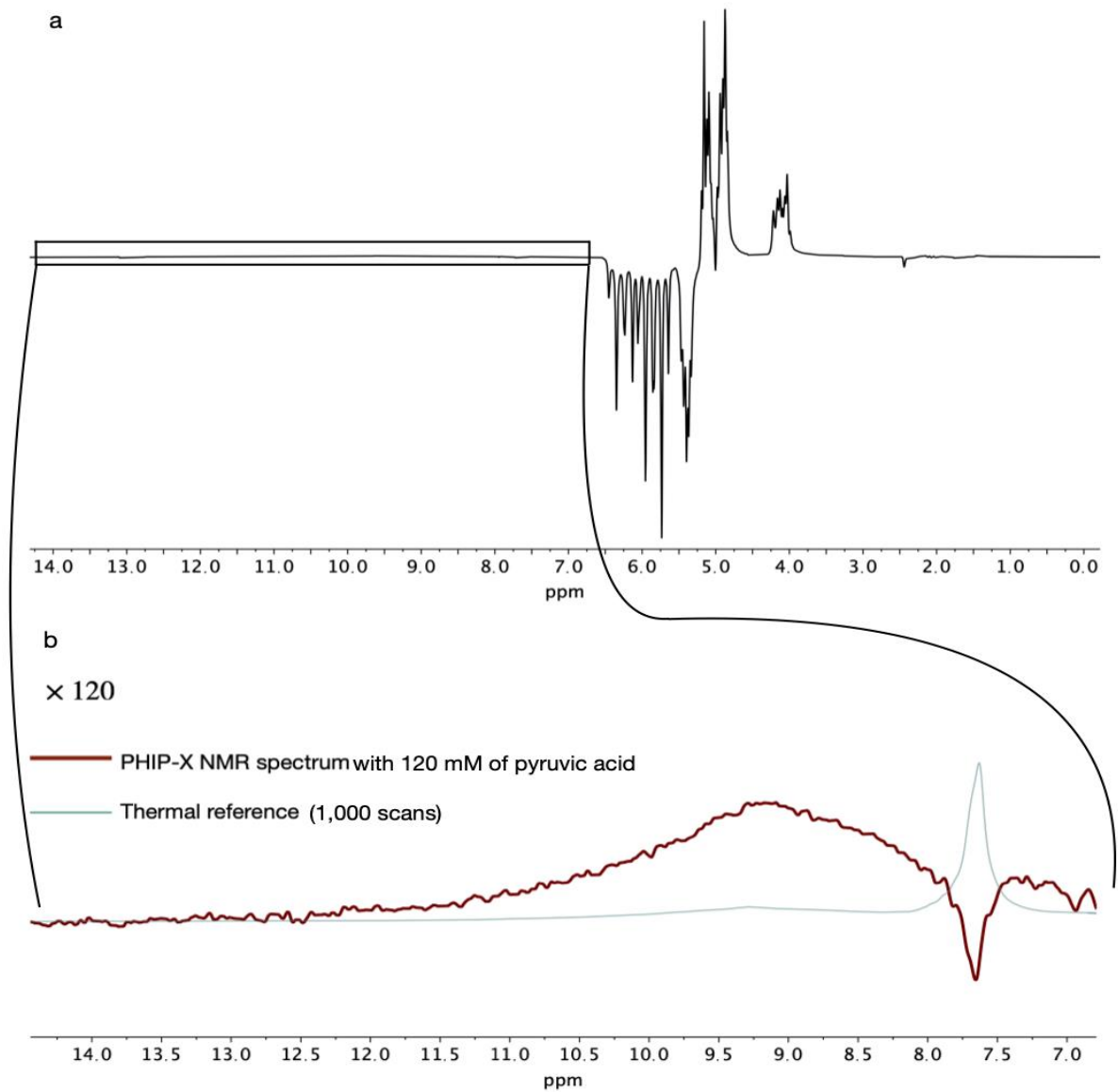

S66: PHIP-X $^1$H NMR spectrum. Parahydrogen with 50% enrichment was bubbled for 20 seconds at atmospheric pressure and 6 mT external field though a solution containing 120 mM of pyruvic acid, 29 mM of propargyl alcohol, and 5.5 mM of [Rh(dppb)(COD)]BF$_4$ in acetone-d$_6$. (a) The full PHIP-X spectrum from 0.0 ppm - 14.0 ppm. Due to the very broad distribution of the COOH peak, a 120-fold amplification in the region from 7.0 ppm - 14-0 is required in order to see NMR signal of the hyperpolarized COOH proton of pyruvic acid. The thermal reference is taken from the same sample used for this PHIP-X experiment and recorded with 1,000 averages. The signal at 7.7 ppm is generated from the catalyst.

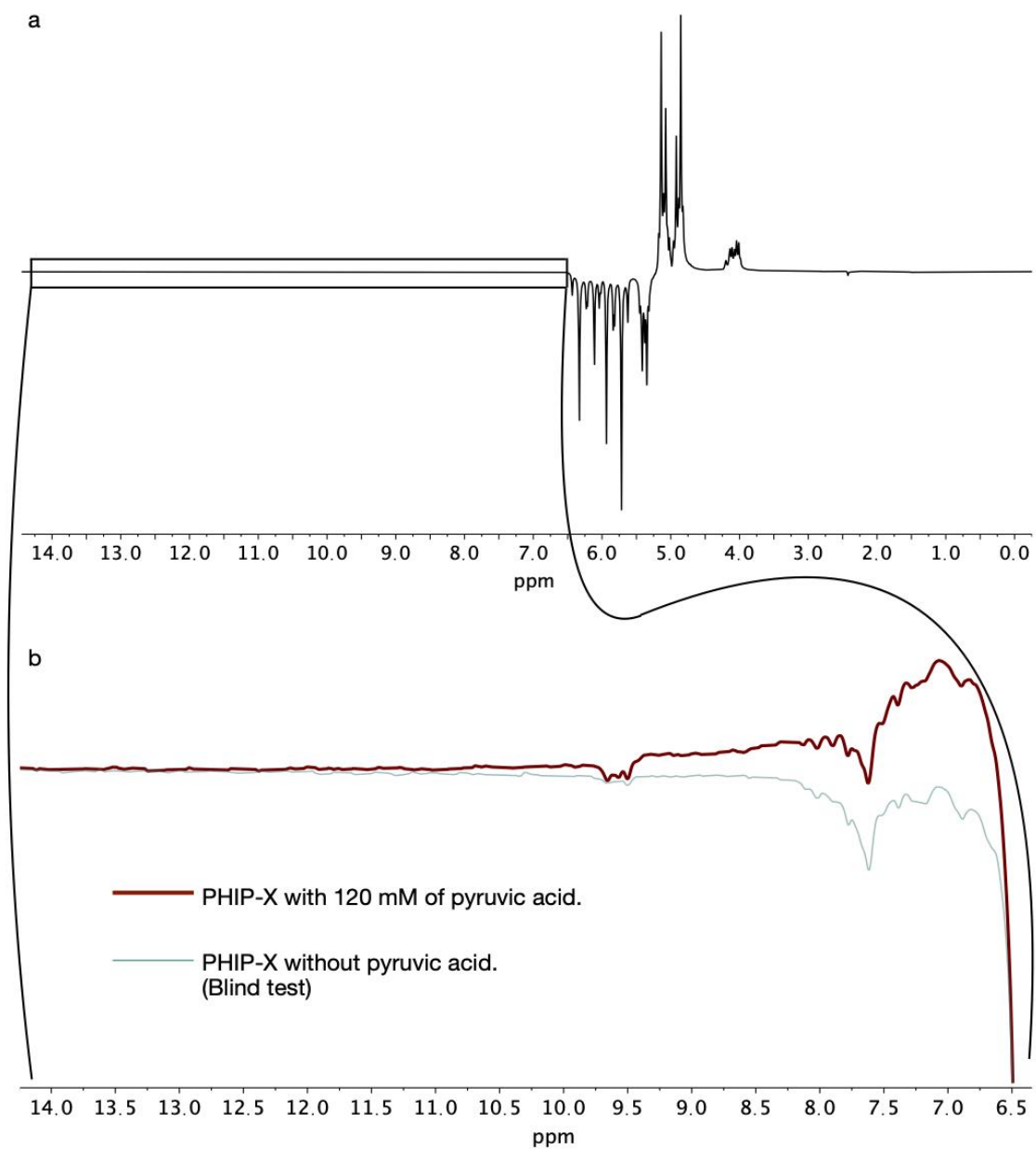

S67: PHIP-X ¹H NMR spectrum with pyruvic acid. $pH_2$ with 50% enrichment was bubbled for 60 seconds at atmospheric pressure and 6 mT external field though a solution containing 29 mM of propargyl alcohol and 5.5 mM of [Rh(dppb)(COD)]BF$_4$ in acetone-d$_6$. Soon after the $pH_2$ supply was stopped, 5 µl of pyruvic acid was injected into the parahydrogenated solution and immediately transferred into the NMR spectrometer. (a) The full PHIP-X spectrum from 0.0 ppm - 14.0 ppm. About 10-fold stronger NMR signals for 1b were detected compared to the case in which pyruvic acid was already present during the hydrogenation reaction. Hence, this approach solves the problem of inhibiting the catalytic parahydrogenation due to the presence of pyruvic acid. However, in this approach the NMR signal of the COOH proton of pyruvic acid becomes strongly distributed and overlaps with signals of 1b. The polarization is very difficult to quantify in this case and may possess an error. However, comparison with a blind test shows a stronger polarization of the COOH proton than when pyruvic acid is present during parahydrogenation of 1a to 1b. A zoom from 6.5 ppm - 14-0 is required in order to see NMR signal of the hyperpolarized COOH proton of pyruvic acid. The blind test used the same parameters and solution as the PHIP-X experiment with pyruvic acid but, of course, pyruvic acid was not injected for the blind test.

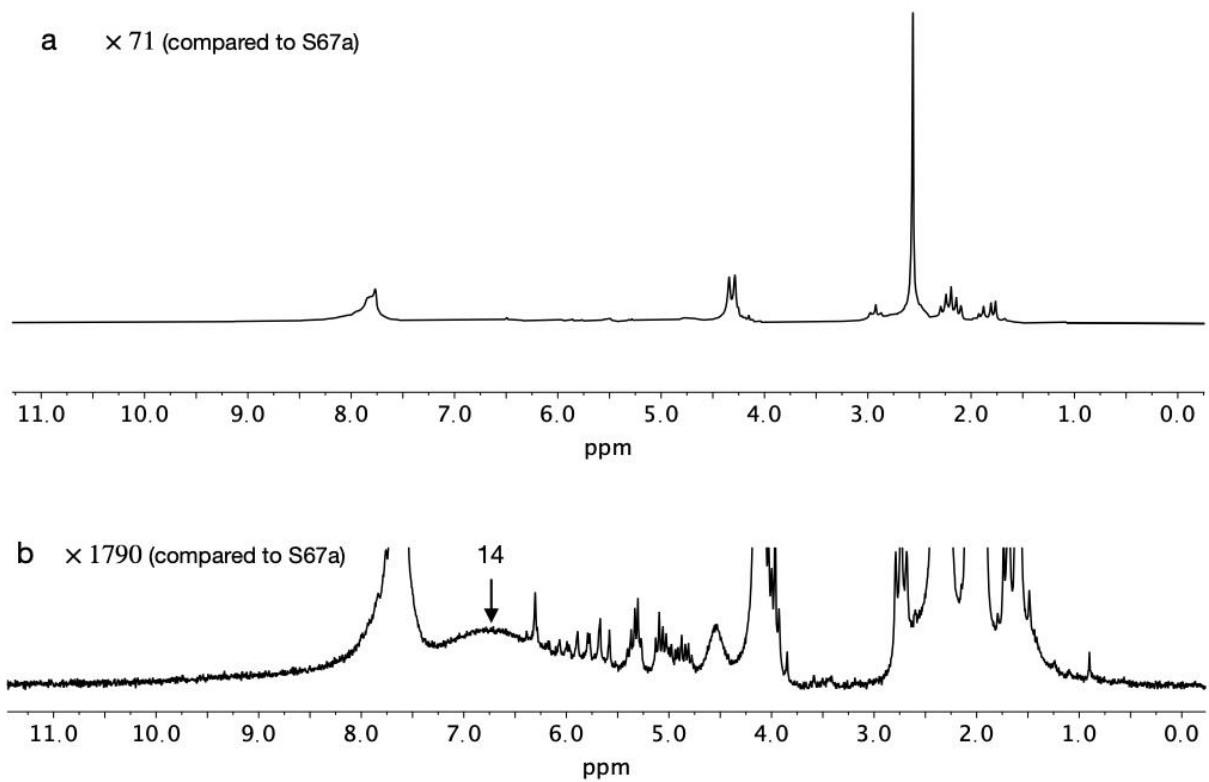

S68: Thermal ¹H NMR spectrum recorded 10 minutes after acquisition of the spectrum shown in fig. S67. Compared to the hyperpolarized spectrum shown in fig. S67 the thermal spectrum (a) is enlarged by a factor of 71. The 1790-fold amplification (b) makes the NMR signal (14) of the COOH proton of pyruvic acid visible. Due to the different chemical environment, this signal is located around 6.75 ppm.

7. Dependence of the PHIP-X polarization on the polarization field.

The PHIP-X polarization transfer to the -$CH_2$- (no. 10 the main manuscript) of the target molecule ethanol was investigated for 11 different polarization fields between the earth's magnetic field and 15 mT. To this end, we have bubbled p$H_2$ with 50% enrichment for 15 seconds at atmospheric pressure and the corresponding polarization field through a solution containing 85 mM propargyl alcohol, 50 mM ethanol, and 5 mM [Rh(dppb)(COD)]$BF_4$ in acetone-$d_6$.

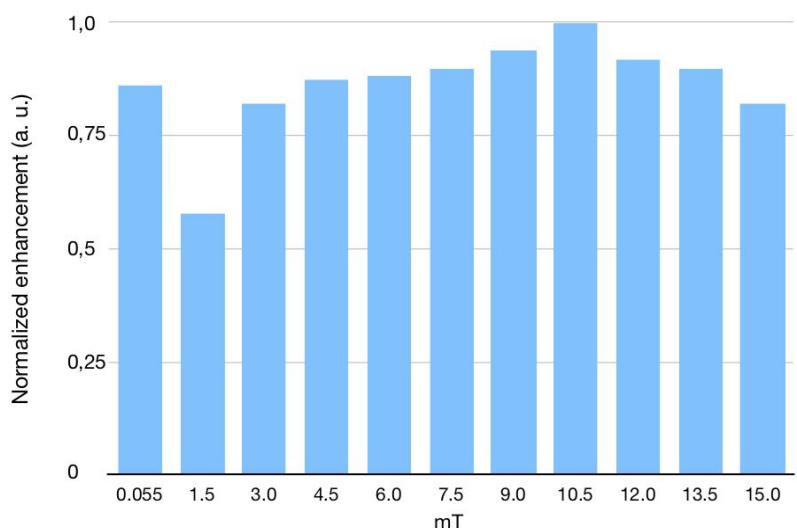

S69: Dependence of the NMR signal enhancement of the -$CH_2$- protons (no. 10 the main manuscript) of the target molecule ethanol on the polarization field. PHIP-X experiments were carried out from 0.055 mT to 15.0 mT. There is a minimum at 1.5 mT and a maximum at 10.5 mT. This indicates the potential to further increase the polarizations reported in the main manuscript.